\documentclass[12pt,preprint]{aastex}
\doublespace
\slugcomment{73 pages, including 12 figures and 7 tables}
\shorttitle{solid surface density of solar nebula}
\shortauthors{Dodson-Robinson et al.}
\begin{document}

\title{Ice Lines, Planetesimal Composition and Solid Surface Density in
the Solar Nebula}

\author{Sarah E. Dodson-Robinson}
\affil{NASA Exoplanet Science Institute, California Institute of
Technology \\
MC 100-22 Pasadena, CA 91125}
\email{sdr@ipac.caltech.edu}

\author{Karen Willacy}
\affil{Jet Propulsion Laboratory,  California Institute of Technology \\
4800 Oak Grove Dr, Pasadena, CA 91109}

\author{Peter Bodenheimer}
\affil{ Department of Astronomy and Astrophysics,  University of
California at Santa Cruz \\
1156 High St, Santa Cruz, CA 95064}

\author{Neal J. Turner}
\affil{Jet Propulsion Laboratory, California Institute of Technology \\
4800 Oak Grove Dr, Pasadena, CA 91109}

\author{Charles A. Beichman}

\affil{NASA Exoplanet Science Institute/Jet Propulsion Laboratory \\
California Institute of Technology \\
MC 100-22, Pasadena, CA 91125}







\begin{abstract}
To date, there is no core accretion simulation that can successfully
account for the formation of Uranus or Neptune within the observed
2-3~Myr lifetimes of protoplanetary disks.  Since solid accretion rate
is directly proportional to the available planetesimal surface density,
one way to speed up planet formation is to take a full accounting of all
the planetesimal-forming solids present in the solar nebula.  By
combining a viscously evolving protostellar disk with a kinetic model of
ice formation, which includes not just water but methane, ammonia, CO
and 54 minor ices, we calculate the solid surface density of a possible
giant planet-forming solar nebula as a function of heliocentric distance
and time.  Our results can be used to provide the starting planetesimal
surface density and evolving solar nebula conditions for core accretion
simulations, or to predict the composition of planetesimals as a
function of radius.

We find three effects that favor giant planet formation by the core
accretion mechanism: (1) a decretion flow that brings mass from the
inner solar nebula to the giant planet-forming region, (2) the fact that
the ammonia and water ice lines should coincide, according to recent lab
results from \cite{collings04}, and (3) the presence of a substantial
amount of methane ice in the trans-Saturnian region.  Our results show
higher solid surface densities than assumed in the core accretion models
of Pollack et al. (1996) by a factor of 3--4 throughout the
trans-Saturnian region.  We also discuss the location of ice lines and
their movement through the solar nebula, and provide new constraints on
the possible initial disk configurations from gravitational stability
arguments.
\end{abstract}


\clearpage

\section{Introduction}
\label{introduction}

Previous core accretion simulations of the formation of Uranus and
Neptune require the solar nebula lifetime to be at least 10~Myr longer
than the observed median protostellar disk lifetime (2--3 Myr; Haisch et
al. 2001).  Similarly, the simulation of Saturn's formation by Alibert
et al. (2001) allows the planet to form in 2.5~Myr, but only if it
migrates from 11.9~AU to 9.5~AU during formation.
This scenario is incompatible with the
outward migration of Saturn predicted by the Nice model of planetary
dynamics (Tsiganis et al. 2005)
This research was motivated by the need to create a model of the
solar nebula with enough solid mass to form Jupiter, Saturn, Uranus and
Neptune via core accretion at heliocentric distances of $5 < R < 20$~AU.

For core accretion simulations, planetesimal mass available for building
giant planet cores is usually calculated by invoking the scaled
minimum-mass solar nebula with surface density $\Sigma \propto
R^{-3/2}$, obtained from spreading out the current mass of the planets
(Weidenschilling 1977), and assuming a gas/solid ratio of $\sim 70$.
However, a single planet core embedded in a disk of planetesimals grows
at the rate of
\begin{equation}
{dM_{\rm core} \over dt} = \pi a_{\rm core}^2 \Sigma_{\rm solid}
\Omega F_g ,
\label{planetgrowth}
\end{equation}
where $F_g$ is a factor taking into account gravitational focusing by
the core (Safronov 1969), $\Sigma_{\rm solid}$ is the solid surface
density of planetesimals, $\Omega$ is the orbital angular frequency of
the core, and $a_{\rm core}$ is the core's effective capture radius.
For Keplerian orbits, we have
\begin{equation}
\Omega = \left ( {GM_* \over R^3} \right )^{1/2} .
\label{kepler}
\end{equation}
The $R^{-3/2}$ dependence of $\Omega$ means that even planets forming
in a disk with uniform surface density should, in the end, reach masses
that suggest $\Sigma \propto R^{-3/2}$ if growth rate is not taken
into account.  A near-uniform surface density distribution, in which
mass is not concentrated in the inner disk, gives higher surface
densities in the outer disk and may speed up the formation
of Uranus and Neptune.

Furthermore, the canonical gas/solid ratio of 70, which is based on the
composition of Comet Halley (Jessberger et al. 1989), includes only one
ice in the solid inventory---H$_2$O.  However, there were certainly
other ices present in the solar nebula: the Deep Impact ejecta from
Tempel 1 contained $\sim 1000$~tons of excavated CO, CO$_2$ and CH$_3$OH
(A'Hearn 2008).  Taking a full accounting of the solids in the solar
nebula may reveal other volatile mass sources besides H$_2$O that
condense in the outer solar nebula and improve the formation prospects
of Uranus and Neptune.

In this work, we calculate the time-evolving solid surface density
available for giant planet formation.  We make two updates to previous
solar nebula simulations: (1) we relax the assumption of a steady-state
disk and allow for viscous redistribution of mass, which tends to drive
the surface density profile toward uniformity, and (2) we add a chemical
reaction network to trace the formation and freezeout of volatiles.

This paper is organized as follows.  In \S \ref{overview}, we outline
our method for calculating solid surface density in the solar nebula:
which codes we use, how they are linked, and how they differ from
previous solar nebula models.  In \S \ref{alphadisk}, we describe our
protostellar disk model and use it to calculate the total surface
density (gas+solid) and temperature evolution of the disk.  \S
\ref{chemmodel} details the physics behind our chemical reaction
network, the abundances of the major ice species, and the location of
the ice lines.  We calculate the solid surface density available for
planet formation in \S \ref{planetesimal} and present our conclusions in
\S \ref{conclusions}.

\section{Simulation Overview}
\label{overview}

This simulation consists of two independent, non-interacting codes used
in sequence.  The first code is a non-steady-state, 1+1-dimensional
$\alpha$-disk model that simulates the dynamical evolution of the solar
nebula.  Code 1 gives the surface density distribution $\Sigma(R,t)$,
along with the vertical and radial temperature and volume density
distributions $T(R,z,t)$ and $\rho(R,z,t)$.

The results from Code 1 fill an important niche in the body of work on
protostellar disk evolution.  Currently available are 2-d, static disk
models that include stellar insolation and radiogenic heating (e.g.
D'Alessio et al. 2006); quasi-steady-state 1+1-d models where
$\dot{M}(t)$ follows an imposed prescription (e.g. Hersant et al. 2001);
and non-steady-state models with only one zone in the vertical direction
(Ruden and Pollack 1991).  The evolving protostellar disks of Boss
(2007) include 3-d hydrodynamics and radiative diffusion, but cover a
timespan of $< 4000$ years.

In this work, we couple full vertical structure models with the
time-dependent, non-steady-state surface density evolution of the disk
and present the first long-term solar nebula simulation with viscous
angular momentum transport determined self-consistently with vertical
structure.  For a description of how the resulting dynamical evolution
of our disk differs from previous simulations, see \S \ref{massflux}.

Code 2 is a non-equilibrium chemical model, including gas reactions,
grain-surface reactions, freezeout and desorption.  This code traces the
abundance of 211 species, including 58 ices, as the disk cools and the
ice lines move inward.  Chemical reaction, freezeout and desorption
rates are governed by the temperature and density of the solar nebula
gas and grains.  We use the midplane temperatures and densities
calculated by Code 1 to generate reaction rate coefficients for Code 2.


Previous chemical models of the solar nebula include those of Aikawa et
al. (1996), Willacy and Langer (2000), and Ilgner and Nelson (2006).
\cite{aikawa96} extended the minimum-mass solar nebula (MMSN; Hayashi
1981) to 800~AU to predict the observed depletion of gaseous CO in the
outer regions of T-Tauri disks.  Their calculation included gas-phase
reactions of CO and other C, H, O, N and S-based molecules, freezeout of
CO and its reaction products, formation of H$_2$, and electron-ion
recombination on grain surfaces.  The density and temperature of the
underlying disk were assumed to stay constant with time, while the
time-evolving midplane CO abundance was calculated with an independent
sequence of radial one-zone models.

Willacy and Langer (2000) calculated the vertical distribution and
column density of common interstellar molecules as a function of time in
the flared protoplanetary disk of Chiang and Goldreich (1997).  They
used a similar species set to Aikawa et al. (1996) but added
grain-surface reactions and photodesorption to the chemistry.  The
simulation consisted of a 2-d grid of noninteracting one-zone models in
the $R-z$ plane.  The underlying temperature and density distribution of
the protostellar disk remained constant with time.

Finally, Ilgner and Nelson (2006) embedded a five-species chemical
reaction network in an evolving $\alpha$-disk model, fully coupling the
chemical and dynamical evolution of the disk.  They used the chemical
reaction network to trace the ionization fraction, which is affected by
charge-transfer reactions, electron-ion recombination, and viscous
diffusion.  The simulation was computationally tractable due to the
small reaction network (5 species, 4 reactions), and the limited extent
of the protostellar disk (0.1--10~AU).

Code 2 of this work is the first to combine an extensive chemical
reaction network, including ice formation and grain-surface reactions,
with a time-evolving protostellar disk model (as used by Ilgner and
Nelson 2006).  Our chemical model consists of a radial sequence of
noninteracting one-zone models, as in Aikawa et al. (1996).  The disk
model in Code 1 therefore functions as a stand-alone input to the
chemical model of Code 2.  Our models differ from those of Aikawa et al.
(1996) and Willacy and Langer (2000) in that we allow the disk
temperature and density to change with time, as prescribed by the
results of Code 1.  Although Code 1 simulates the full vertical
structure of the protostellar disk, we apply Code 2 only to the
midplane, which is where solids settle and build planetesimals.

Code 2 calculates the abundance of each ice and gas species as a
function of heliocentric radius and time.  Combining these results with
information from the literature about the composition of refractory
grains, we calculate the gas/solid mass ratio in the solar nebula.  This
ratio tells us what fraction of the total surface density calculated by
Code 1 is solid and can build giant planet cores.  The end result of
both codes is the solid surface density distribution $\Sigma_{\rm
solid}(R,t)$.

Since most molecules in the solar nebula formed in a molecular cloud, we
first run Code 2 for molecular cloud conditions ($T = 10$~K, hydrogen
number density $2 \times 10^4 \; {\rm cm}^{-3}$, grain radius $0.258 \mu
m$).  The molecular cloud model begins with atomic gas of solar
composition (except for hydrogen, 99\% of which is in H$_2$) and bare
refractory grains at the standard interstellar abundance $n_g = 10^{-12}
(n_{\rm H} + 2 n_{\rm H2})$, where $n_g$ is the grain abundance and
$n_{\rm H}$ and $n_{\rm H2}$ are the abundances of H and H$_2$,
respectively.  We run the molecular cloud model for $10^6$~yr and use
the resulting ice-gas mixture as the starting composition of the solar
nebula.

\section{Disk Model}
\label{alphadisk}

\subsection{Overall Model Formulation}
\label{framework}

The goals of this simulation are to calculate the time-evolving
temperature and density in the solar nebula, which determine the
reaction rates for our kinetic model of ice formation, and to follow the
viscous evolution of the surface density profile.  Although from a
planet formation perspective we are primarily concerned with the ice
inventory at the solar nebula midplane, an accurate dynamical
description of the midplane requires understanding the vertical energy
balance in the disk.  To simplify the problem of long-term protostellar
disk evolution, we use the following assumptions:

\begin{enumerate}

\item The disk is axisymmetric and symmetric about the midplane.

\item The disk is geometrically thin: $H / R << 1$.

\end{enumerate}

The disk is represented in cylindrical polar coordinates $R$
(heliocentric distance) and $z$ (height above midplane), where the
$z$-axis is the rotation axis of the nebula.  Assumption 1 reduces the
physical 3-dimensional disk to a 2-dimensional quadrant with zero flux
at the midplane, allowing us to suppress the azimuthal coordinate
$\theta$.  Assumption 2 allows the vertical and radial dimensions of the
disk to be decoupled to form a 1+1-dimensional framework: at each radial
gridpoint resides an independent vertical structure model.  At each
timestep, energy balance between viscous heating and radiative cooling
is solved within in a single annulus.

\subsection{Radial Diffusion}
\label{raddiff}

We calculate the radial motion of mass using the surface density
diffusion equation of Lynden-Bell and Pringle (1974):
\begin{equation}
{\partial \Sigma \over \partial t} = {3 \over R} {\partial \over
\partial R} \left [ R^{1/2} {\partial \over \partial R} \left ( \Sigma
\nu R^{1/2} \right ) \right ] .
\label{diffusion}
\end{equation}
Eq. \ref{diffusion} is valid when the protostar's accretion rate is
small, so the sun is almost fully assembled at the beginning of the
simulation.  This assumption is consistent with the T-Tauri phase of
disk evolution.  We take the viscosity $\nu$ governing radial diffusion
to be the midplane value.

To calculate $\nu_{\rm midplane}$, we make the following 
assumptions about angular momentum transport in the disk:

\begin{enumerate}

\item {\it Viscous stresses follow the standard $\alpha$-viscosity
prescription} (Shakura and Syunyaev 1973).  We assume the
magnetorotational instability (Balbus and Hawley 1991; hereafter MRI)
provides the requisite turbulent viscosity.

\item {\it The grains and gas are well mixed.} This assumption allows us
to track the disk's solid surface density evolution by modeling only a
single fluid.  

\end{enumerate}

The turbulent viscosity is
\begin{equation}
\nu = \frac{2}{3} \alpha c_s H , 
\label{visc}
\end{equation}
where $c_s$ is the isothermal sound speed and $H$ is the modified
pressure scale height, softened into a nonsingular form (Milsom et al.
1994):
\begin{equation}
H = { c_s / \Omega \over \sqrt{ 1 + \left ( {2
z^2 \Omega^2 / c_s^2} \right ) }} .
\label{scale}
\end{equation}
The angular velocity, neglecting disk self-gravity, is
\begin{equation}
\Omega = \left [ {G M_* \over \left ( R^2 + z^2
\right )^{3/2} } \right ]^{1/2} ,
\label{omega}
\end{equation}
and the isothermal sound speed is
\begin{equation}
c_s^2 = {R_g \over \mu} T .
\label{soundspeed}
\end{equation}
In parts of the disk where $H > z_{\rm surf}$, the disk surface height,
we substitute $z_{\rm surf}$ for $H$ in Eq. \ref{visc}.  We take a
mean molecular weight of $\mu = 2.33$~g~mol$^{-1}$ (Ruden and Pollack
1991).  Based on recent global-disk MHD simulations (Lyra et al. 2008),
we choose $\alpha = 2 \times 10^{-3}$.  We further discuss the interplay
between $\alpha$, disk mass and surface density profile, along with our
constraints on these parameters, in \S \ref{initial}.  All free
parameters in the disk model are listed in Table \ref{freedisk}.


Although we assume turbulence is caused by MRI, we do not include the
radial variation of the viscosity coefficient $\alpha$ caused by the
dead zone (e.g. Kretke and Lin 2007, Reyes-Ruiz et al. 2003).
However, we do account for the cessation of MRI
turbulence in the tenuous outer regions of the disk, where ions and
neutrals may be only weakly coupled.  \cite{hawley98} find that neutrals
participate fully in MRI turbulence only when the ion-neutral collision
timescale $T_{\rm coll}$ is much less than the orbital timescale;
$T_{\rm coll} \leq 0.01 / \Omega$.  Ions and neutrals decouple and MRI
turbulence ceases when $T_{\rm coll} \geq 100 / \Omega$.  In the
transition region between coupled and decoupled neutral-ion interaction,
where $0.01 \leq T_{\rm coll} \Omega \leq 100$, we decrease $\log
\alpha$ linearly from its fiducial value.

To calculate $T_{\rm coll}$, we find the mean free time of a neutral
particle traveling through a gas of moving ions.  The neutrals and ions
collide at an average angle of $90^{\circ}$ and thus a mean impact
velocity of
\begin{equation}
v_{\rm coll} = \sqrt{v_n^2 + v_i^2} = v_n \left (1 + {\mu_n \over \mu_i}
\right )^{1/2} ,
\label{vcoll}
\end{equation}
where $\mu_n$ and $\mu_i$ are the molecular weights of neutrals and
ions, respectively.  The mean collision time is then
\begin{equation}
T_{\rm coll} = {1 \over \pi \left ( a_n + a_i \right )^2 n_i v_n
\left (1 + \mu_n / \mu_i \right )^{1/2}} ,
\label{tcoll}
\end{equation}
where $a_n$ and $a_i$ are the radii of neutrals and ions, $v_n =
\sqrt{3} c_s$ is the mean velocity of a neutral and $n_i$ is the number
density of ions:
\begin{equation}
n_i = {f_i \rho \over m_i}
\label{ni}
\end{equation}
We assume an ionization fraction of $f_i = 10^{-9}$ and an ion mass of
$m_i = 30$~AMU.  $f_i$ and $m_i$ are listed along with other free
parameters in the simulation in Table \ref{freedisk}.

\subsection{Vertical Structure}

Since the disk midplane viscosity is determined by the balance between
viscous energy generation and radiative energy losses at the
photosphere, we calculate full vertical structure models at each
radial grid point.  We use the following assumptions about the mass and
energy balance in the vertical direction:

\begin{enumerate}

\item {\it The disk stays in hydrostatic equilibrium}.  Even though the
mass distribution evolves, the vertical sound crossing timescale is much
less than the radial diffusion timescale.

\item {\it We neglect stellar irradiation as an energy source, as it has
little effect on the midplane temperature}.  In section \ref{initial} we
show that the disk is self-shadowed ($H/R$ decreases with $R$), making
this assumption self-consistent with our simulation results.

\end{enumerate}

A system of three coupled differential equations specifies the vertical
structure of the disk.  First, the disk is in hydrostatic equilibrium:
\begin{equation}
{\partial P \over \partial z} = -\rho \Omega^2 z .
\label{hydrostatic}
\end{equation}
Second, the viscous energy generation rate per unit volume (Pringle
1981) is
\begin{equation}
{\partial F \over \partial z} = \frac{9}{4} \nu \Omega^2 \rho .
\label{flux}
\end{equation}
Third, where vertical energy transport is regulated by radiative
diffusion and energy transport in the radial direction is negligible
(thin disk assumption), the energy equation is
\begin{equation}
{\partial T \over \partial z} = {\partial P \over \partial z} \nabla {T
\over P} .
\label{tempgrad}
\end{equation}

To calculate the thermodynamic gradient $\nabla = d \ln T / d \ln P$, we
use the Schwarzschild criterion for stability against convection
(Kippenhahn and Weigert 1994):
\begin{equation}
\nabla = \left \{ \begin{array}{l} \nabla_{\rm rad},  
\nabla_{\rm rad} \le \nabla_{\rm ad} \\ \nabla_{\rm conv},  
\nabla_{\rm rad} > \nabla_{\rm ad} \end{array} \right. .
\label{nabla}
\end{equation}
In Eq. \ref{nabla}, $\nabla_{\rm ad}$ is the adiabatic gradient,
which is $2/7$ for a diatomic gas; $\nabla_{\rm conv}$ is the convective
gradient; and $\nabla_{\rm rad}$ is the radiative gradient: 
\begin{equation}
\nabla_{\rm rad} = \frac{3}{4} {\kappa P F \over a c \Omega^2 z T^4} ,
\label{nablarad}
\end{equation}
where $a$ is the radiation density constant and $\kappa$ is the local
Rosseland mean opacity (for a description of the opacity tables used in
this calculation, see \S \ref{opsec}).

We use mixing-length theory to calculate $\nabla_{\rm conv}$.  We begin
the convection treatment by defining two dimensionless quantities:
\begin{equation}
W \equiv \nabla_{\rm rad} - \nabla_{\rm ad}
\label{convw}
\end{equation}
and
\begin{equation}
U \equiv {3 a c T^3 \over c_P \rho^2 \kappa \ell_m^2}
\sqrt{8 \varepsilon} .
\label{convu}
\end{equation}
In Equation \ref{convu}, $c_P$ is the specific heat at constant pressure
and $\ell_m$ is the mixing length.  As in \cite{milsom94}, we define
$\ell_m$ as the minimum of the convection zone top or the pressure scale
height $H$ (Equation \ref{scale}).  $\varepsilon$ is an analytical
softening of the ratio $H / \Omega^2 z$, eliminating the singularity at
$z = 0$:
\begin{equation}
\varepsilon = \sqrt{1 \over \Omega^4 (1 + z^2 / H^2)} .
\label{varparm}
\end{equation}
Defining a new variable $\xi$ as
\begin{equation}
\xi \equiv \nabla_{\rm conv} - \nabla_{\rm ad} + U^2 = \sqrt{\nabla_{\rm
conv} - \nabla_e} + U
\label{xieqn}
\end{equation}
(where $\nabla_e$ is the thermodynamic gradient of an individual
convective element), we can then solve the cubic equation
\begin{equation}
\left ( \xi - U \right )^3 + \frac{8U}{9} \left ( \xi^2 - U^2 - W \right
) = 0
\label{cubic}
\end{equation}
to find $\nabla_{\rm conv}$.  For the derivations of Equations
\ref{xieqn} and \ref{cubic} see Chapter 7 of \cite{kippenhahn94}.

Finally, we close the system of equations with the ideal gas equation of
state,
\begin{equation}
P = \left ({R_g \over \mu} \right ) \rho T .
\label{idealgas}
\end{equation}

\subsection{Computational Methods and Boundary Conditions}

We begin the simulation with a specified surface density profile for the
disk, $\Sigma \propto r^{-3/2}$ (Hayashi 1981).  The total disk mass
used to normalize $\Sigma(r)$ and the viscosity coefficient $\alpha$ are
determined by constraints of gravitational stability
and the need for enough mass to form
giant planets.  For full details of how the $t = 0$ disk is constructed,
see \S \ref{initial}.  We also assume the disk is embedded in a medium
with an ambient temperature $T_{\rm amb}$ (for example, a molecular
cloud core).  This places a firm lower limit on the nebula temperature
and forces inactive regions of the disk to assume an isothermal vertical
structure.

As in \cite{bell97}, we place the disk surface at
optical depth $\tau = 0.03$.  We begin the vertical structure
calculation with guesses for the free parameters $T_{\rm surf}$ and
$\rho_{\rm surf}$, the density and temperature at the $\tau = 0.03$
surface.

The net flux leaving the disk surface is determined by the accretion
temperature $T_{\rm acc}$, generated solely by viscous stresses:
\begin{equation}
F = \sigma T_{\rm acc}^4 .
\label{teff}
\end{equation}
We require a value of $T_{\rm acc}$ at the disk surface to to start
integrating Equations \ref{flux} and \ref{tempgrad}.  The free parameter
$T_{\rm surf}$ is the flux sum of the accretion temperature $T_{\rm
acc}$ and the ambient temperature $T_{\rm amb}$, invoking the Eddington
approximation:
\begin{equation}
T_{\rm surf}^4 = T_{\rm amb}^4 + \frac{3}{4} \left [ \tau + f \left (
\tau \right ) \right ] T_{\rm acc}^4 .
\label{fluxsum}
\end{equation}
In equation \ref{fluxsum}, $f(\tau)$ is the Hopf function, with a value
of 0.601242 for $\tau = 0.03$ (Bell et al. 1997; Mihalas 1978).

After calculating $P$ from Eq. \ref{idealgas}, the variables of
integration $T_{\rm acc}$, $P$ and $F$ are all specified.  We find the
height of the $\tau = 0.03$ surface by solving the following equation
for z:
\begin{equation}
{ 1 \over \Omega^2 \, z } = {\tau \over \kappa(\rho, T) \, P} .
\label{zsurfeqn}
\end{equation}
Integration proceeds inward from the disk surface to the midplane using
the Runge-Kutta method with adaptive stepsize control (Press et al.
1992).  We repeat the integration, adjusting $T_{\rm surf}$ and
$\rho_{\rm surf}$ with the Newton-Raphson algorithm (Press et al. 1992),
until a solution is found that satisfies
\begin{equation}
F_{\rm midplane} = 0
\label{fluxconv}
\end{equation}
(following Assumption 1, \S \ref{framework}) and
\begin{equation}
2 \int_{z=0}^{z_{\rm surf}} \rho \: dz = \Sigma .
\label{sdconv}
\end{equation}

After the vertical structure is solved for each radius, we use the
resulting $\nu_{\rm midplane}$ to update $\Sigma(r)$.  We implement
implicit finite differencing to solve Eq. \ref{diffusion},
adjusting the timestep $\Delta t$ so that surface density varies by a
maximum of 1.5\% during a single timestep (e.g. Ruden and Pollack 1991).

The radial grid is equispaced in $r^{1/2}$ and contains 600 cells
between the inner boundary at $0.3$~AU and the initial outer boundary at
$30$~AU.  We implement a zero viscous-stress inner boundary condition,
so matter falls directly onto the star from the innermost gridpoint.
Although the actual accretion shock is at a much smaller radius than
$0.3$~AU, \cite{ruden86} find that the location of the inner boundary
$R_{\rm in}$ is unimportant as long as $R_{\rm out} >> R_{\rm in}$.  We
allow the disk to expand freely from the original outer boundary, so new
radial grid cells are added to the disk as the simulation progresses.

\subsection{Opacities}
\label{opsec}

Solving Eq. \ref{nablarad} requires knowing local mean opacity.
Semenov et al. (2003) calculated Rosseland mean opacities due to icy
grains for temperatures down to 5K.  The authors modeled several
different grain types, including spheres, aggregates, and porous grains.
We use the 5-layered sphere topology, where each grain consists of a
silicate/iron nucleus covered with successive layers of volatiles as the
temperature decreases.  Following Semenov et al., we assume the grains
condensed from gas of solar composition, with the abundances given by
Helling et al. (2000).  These abundances are primarily derived from the
catalog of Anders and Grevesse (1989).  We list the abundances of the
most common elements in Table \ref{solarcomp}, along with the types of
solids they can form: rock, metal, ice, or refractory CHON (e.g.
graphite or kerogen; see Jessberger et al. 1988).

At temperatures above 1200K, iron and silicates sublimate and the main
opacity source switches from dust to molecular gas.  Semenov et al.
(2003) caution that the gas opacities in their model are only
approximate in the temperature range where dust and gas opacity are
comparable.  For temperatures above 1000K, we use the opacities of
Ferguson et al. (2005), updated for our chosen set of solar abundances
(J. Ferguson, private communication).  It should be noted that the
number of atomic and molecular lines included in the calculation of an
opacity table has a greater impact on the resulting opacities than minor
changes in assumed solar composition.

\begin{figure}
\centering
\includegraphics[scale=0.9]{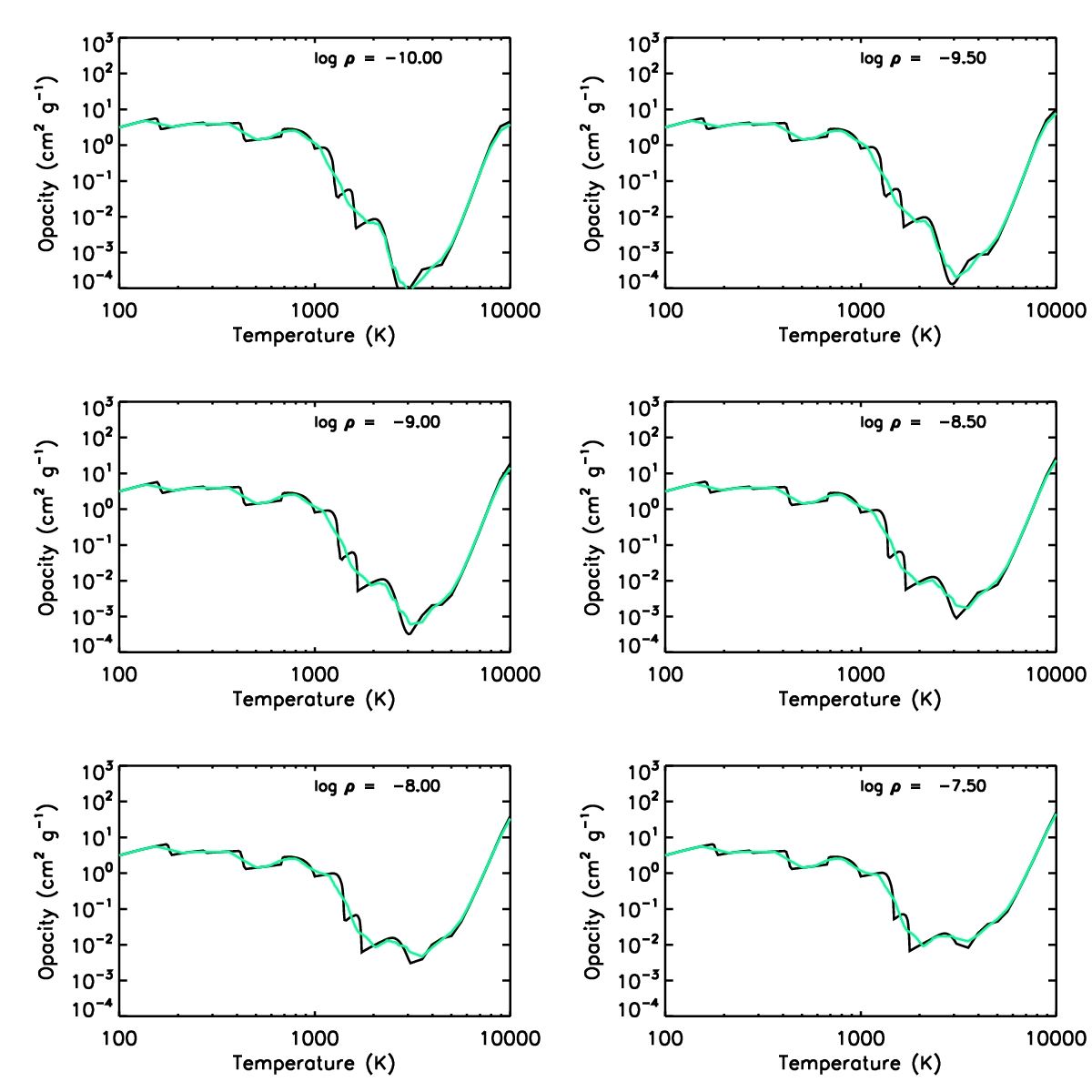}
\caption{Opacity as a function of temperature and density.  The black
line shows the unsmoothed tables of Ferguson et al. (2005) ($T >
1000$K)and Semenov et al. (2003) ($T < 700$K).  In the range $700 < T <
1000$K, we take a weighted average in log-T space of the two tables.
Finally, the green curve shows the smoothed opacity functions we use in
the model.}
\label{opplot}
\end{figure}

Regions where opacity is a strong function of temperature can cause
convergence problems for the Newton-Raphson algorithm, which will
oscillate between two solutions with similar temperatures but quite
different opacities.  To mitigate this problem, we smoothed the raw
opacity tables with an averaging filter.  Figure \ref{opplot} shows
opacity as a function of temperature and density in our model.

\subsection{Initial Conditions}
\label{initial}

Constraints on the early stages of the solar nebula come from
observations of Class 1 protostars and T-Tauri stars and the presence
and orbital configuration of the planets.  We require our $t = 0$ disk
to meet two criteria:


\begin{enumerate}

\item The disk is gravitationally stable to axisymmetric perturbations
at all radii.


\item The disk is massive enough to produce Jupiter and Saturn by core
accretion within the observed mean protostellar disk lifetime of
2--3~Myr (Haisch et al. 2001).

\end{enumerate}

To test whether a particular disk meets Criterion 1, we calculate the
Toomre Q parameter at every radial gridpoint.  Gravitational stability
requires that
\begin{equation}
Q = {c_s \Omega \over \pi G \Sigma} > 1 .
\label{toomreq}
\end{equation}
Gravitationally unstable disks will fragment and form stellar or
substellar clumps on a $10^3$ year timescale.  The lack of a binary
companion for the sun and the near-circular, non-chaotic orbits
displayed by all the planets makes gravitational fragmentation anytime
during solar nebula evolution unlikely.

Our $Q > 1$ criterion determines the combinations of viscosity parameter
$\alpha$, mass distribution and total disk mass that create viable
accretion disks.  High values of $\alpha$ act to stabilize the disk by
increasing energy generation and therefore sound speed.  Large disk
masses destabilize the disk by increasing $\Sigma$.  Surface density
profiles that decline steeply with radius damp gravitational instability
by loading the mass preferentially toward the inner disk, where
dissipative shear stresses are strongest.

Hersant et al. (2001; hereafter HGH) used an evolving $\alpha$-disk
model and measured solar-system D/H ratios to constrain the structure of
the solar nebula.  They chose the surface density profile $\Sigma
\propto R^{-1}$.  Even with an extremely high turbulent energy
generation rate, $\alpha \geq 0.01$, disks with this surface density
distribution are gravitationally stable only if $M_{\rm disk} \leq 0.08
M_{\odot}$.  However, Criterion 3 requires that there be sufficient
solid mass to form Jupiter and Saturn within $\sim 3$~Myr.  Assuming a
gas/solid mass ratio of 70 (Pollack et al. 1996), a disk with $\Sigma
\propto R^{-1}$ out to 30~AU and $M_{\rm disk} = 0.08 M_{\odot}$ has a
solid surface density of only 6~g~cm$^{-2}$ in Saturn's feeding zone at
9.5~AU.  \cite{hubickyj05} found that a planet forming from
6~g~cm$^{-2}$ of solids at 5~AU takes more than 13~Myr to initiate rapid
gas accretion, necessary for forming a massive atmosphere.  Since planet
formation proceeds more slowly at larger heliocentric distances (Eq.
\ref{planetgrowth}), this disk is unlikely to be able to form Saturn.
We thus require a more massive disk, with a steeper surface density
profile to keep the outer regions gravitationally stable.

We choose a disk mass $M_{\rm disk} = 0.12 M_{\odot}$ and a surface
density $\Sigma \propto R^{-3/2}$ (Hayashi 1981).  We take $\alpha = 2
\times 10^{-3}$, the approximate azimuthally and vertically averaged
value at Jupiter's heliocentric distance in the global MHD turbulence
simulations of Lyra et al. (2008).  Our solar nebula has inner and outer
radii of $R_{\rm in} = 0.3$ AU and $R_{\rm out} = 30$~AU, covering the
range of heliocentric distances where planets exist today.  It is worth
noting that studies of the relative motions of dust and gas predict a
much larger outer radius: the gas drag on cm-sized bodies can cause them
to be quickly lost to the Sun unless they have a very large distance to
migrate (Stepinski and Valageas [1997]; Ciesla and Cuzzi [2006]).
Mitigating the deleterious effect of gas drag on large grains may
require that the nebula has an additional low-density component outside
$R = 30$~AU, which we do not model in this work.

We assume our $t = 0$ disk coincides with the beginning of the sun's
T-Tauri phase.  Most of the sun's mass is already in place: $M_* = 0.95
M_{\odot}$ at the start of our simulation.  We assume an ambient
temperature of 20~K, approximating a disk embedded in the remnants of a
molecular cloud.  All free parameters in the disk model are listed in
Table \ref{freedisk}.

Although our model neglects solar irradiation, we find that the giant
planet-forming region of our t=0 disk is in fact shadowed.  The maximum
aspect ratio $H / R$ of the $\tau = 2/3$ photosphere occurs at $R = 2$
AU.  Beyond this radius, no part of the disk receives direct sunlight.
The shadow persists until $t \sim 1$~Myr, when regions with $R > 20$~AU
attain an aspect ratio greater than the local maximum at 2~AU.  In this
way, our disk evolution is broadly consistent with model T-Tauri disk
SEDs, which require flaring to explain observed spectra (e.g. Chiang et
al. 2001), yet the giant planet-forming region ($\sim 5$--20~AU) still
receives no solar heating during the icy planetesimal formation epoch.

\subsection{Disk Model Results}
\label{viscous}

\subsubsection{Surface Density}
\label{surfacedensity}

\begin{figure}
\centering
\includegraphics[scale=0.9]{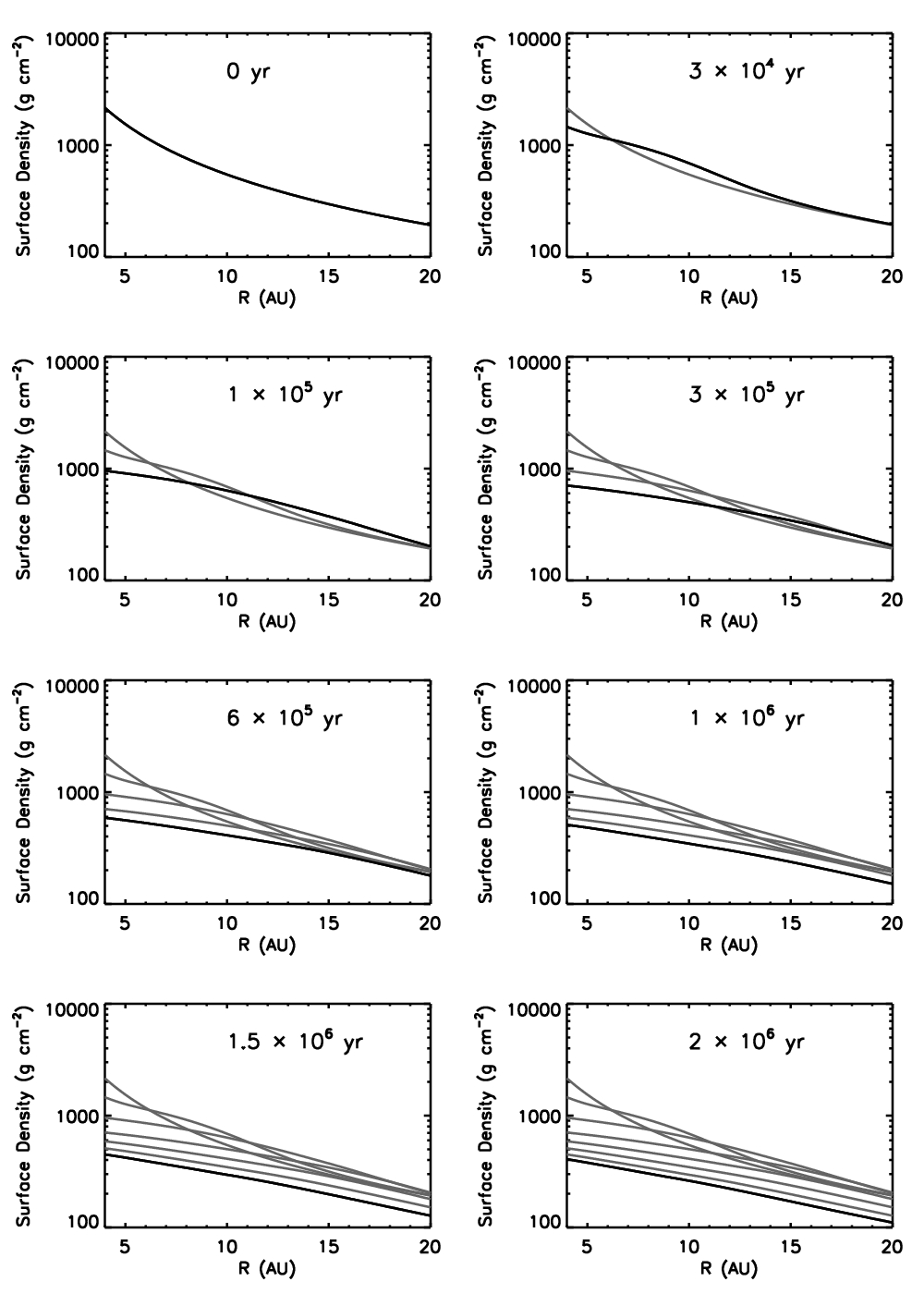}
\caption{Surface density as a function of radius and time in the giant
planet-forming region.  Each panel shows a snapshot of the disk surface
density profile in black, with the profiles from previous epochs
retained in gray.  During the first $5 \times 10^5$~yr of evolution, a
wave of local surface density enhancement propagates outward, depositing
mass in the giant planet region of the solar nebula.}
\label{sdplanet}
\end{figure}

\begin{figure}
\centering
\includegraphics[scale=0.9]{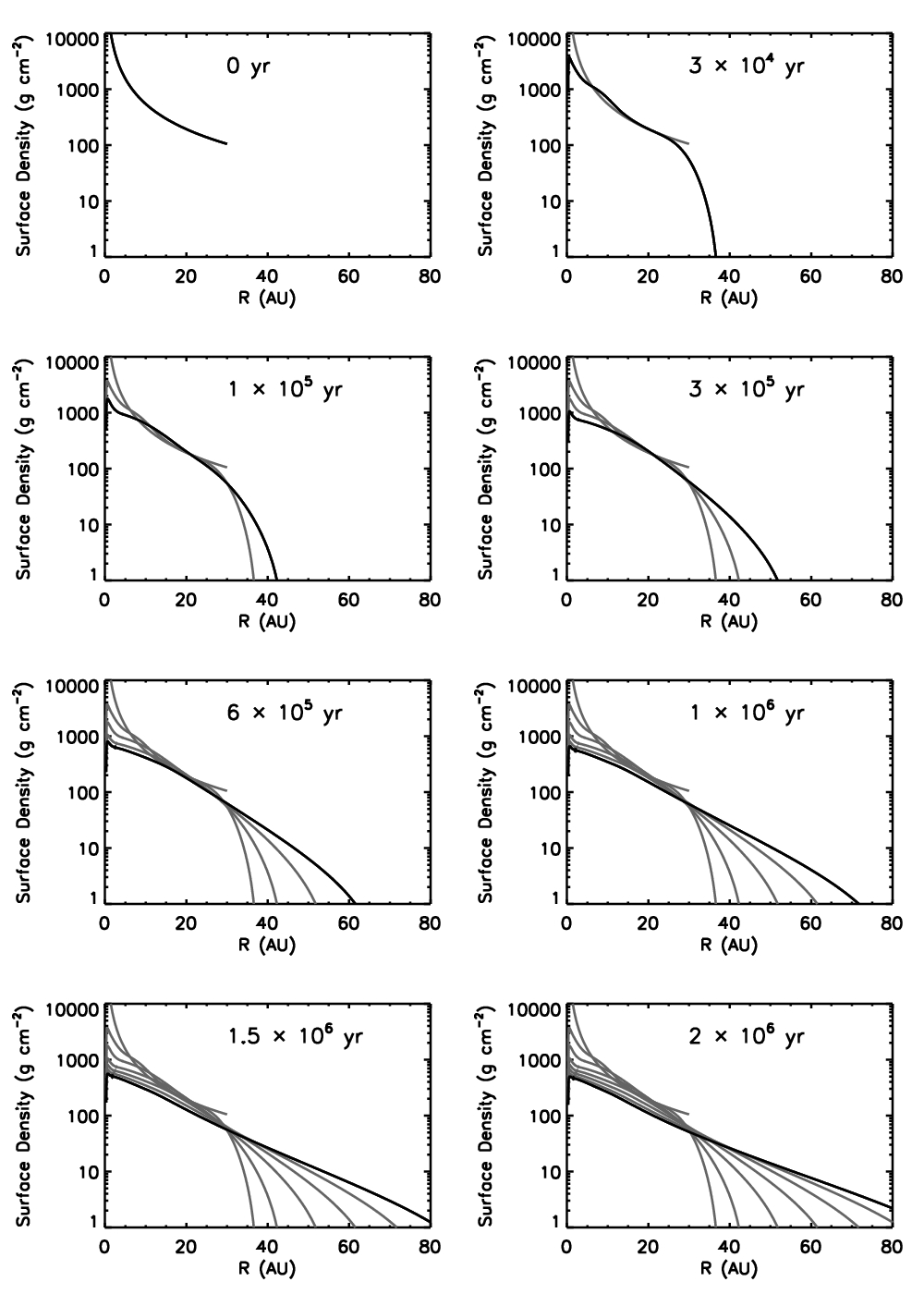}
\caption{Surface density as a function of radius and time throughout the
whole disk.  Each panel shows a snapshot of the disk surface density
profile in black, with the profiles from previous epochs retained in
gray.  The disk expands from $30$~AU to $> 80$~AU over the course of the
simulation.}
\label{sdwholedisk}
\end{figure}

We run the solar nebula model for 2~Myr, about the median age of
observed protostellar disks (Haisch et al. 2001).  Figure \ref{sdplanet}
shows the evolution of the disk surface density profile in the giant
planet-forming region, 5-20~AU (Tsiganis et al. 2005).  The surface
density profile at $t = 0$ is shown on the upper left.  Subsequent
panels show snapshots of $\Sigma(R)$ as the simulation progresses
plotted in black.  For reference, surface density profiles from
preceding plot panels are retained in gray.  Figure \ref{sdwholedisk}
shows the surface density evolution of the entire disk as it expands
from 30~AU at $t = 0$ to $> 80$~AU after 2~Myr.

From examining Fig. \ref{sdplanet}, we can divide the evolution of the
solar nebula's giant planet-forming region into two epochs.  In Epoch 1,
the surface density profile flattens as mass is redistributed from the
inner disk to the outer disk.  At $t = 0$, the net mass flow is outward
everywhere except at $R < 1$~AU, which is accreting inward toward the
star.  A wave of local surface density enhancement consequently
propagates outward, reaching $\sim 10$~AU at $t = 3 \times 10^4$ yr and
dissipating near $\sim 17$~AU at $t = 3 \times 10^5$ yr.  At the end of
Epoch 1, at $t \sim 5 \times 10^5$~yr, the solar nebula has assumed a
surface density profile flatter than $\Sigma \propto R^{-1}$ (equal mass
in each annulus) in all but the inner $0.5$~AU of the disk.  The effect
of this mass redistribution epoch, therefore, is to drive the surface
density distribution toward uniformity.

In Epoch 2, $t > 5 \times 10^5$~yr, surface density decreases with time
everywhere in the disk.  In Fig. \ref{sdwholedisk}, we see the solar
nebula grow more tenuous as it expands: after 2 Myr, the total disk mass
is 0.067~$M_{\odot}$ spread over 80~AU, as opposed to 0.12~$M_{\odot}$
within 30 AU at the beginning of the simulation.  The protosun reaches a
final mass of 1.003~$M_{\odot}$.

\begin{figure}
\centering
\includegraphics[scale=0.9]{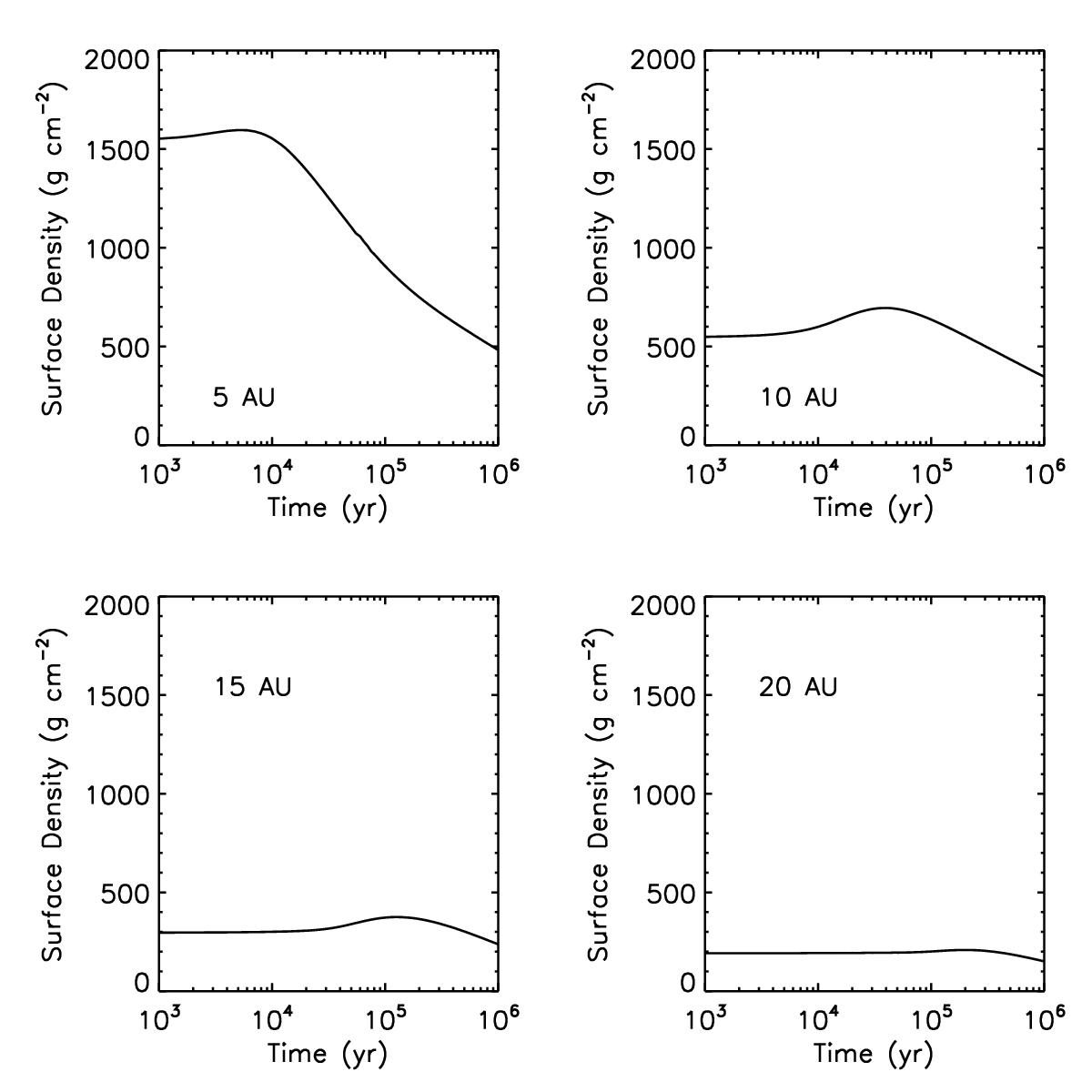}
\caption{Surface density as a function of time at 5, 10, 15 and 20~AU.
An outward-propagating wave adds mass to each annulus as it passes
through.  The largest relative surface density increase, $\Delta \Sigma
/ Sigma_0$, occurs at 10~AU, near Saturn's feeding zone.}
\label{sdann}
\end{figure}

Figure \ref{sdann} shows surface density as a function of time at 5~AU,
10~AU, 15~AU and 20~AU.  At each heliocentric distance, we see the
temporary mass growth as the density wave passes through.  The largest
relative surface density increase $\Delta \Sigma / \Sigma_0$ occurs at
11.5~AU, where the initial surface density of 442~g~cm$^{-2}$ grows by
28\% to 558~g~cm$^{-2}$ in 85,000~yr.  {\it This surface density
enhancement moving outward through the giant planet-forming region
strongly favors solid planet-core formation.}

At the beginning of the sun's T-Tauri phase, before substantial grain
growth occurs, gas and grains are well mixed: accreting gas drags grains
with it.  If gas moves inward with a constant mass flux, surface density
decreases with time everywhere in the disk and solids are lost into the
sun.  Our net outward gas flow deposits small grains in the 5--15~AU
region within the planetesimal formation timescale of $\sim 5 \times
10^4$~yr (Hubbard and Blackman 2006).  {\it These grains provide surface
area for ice deposition and add solid mass to the giant planet feeding
zones, speeding up solid core formation.}

The existence of Epoch 1 and consequent driving of the surface density
distribution toward uniformity is a robust result that we see even using
$t = 0$ surface density profiles with power-law indices of 0.7 and 1.
Our chosen fiducial disk remains stable to axisymmetric perturbations
(c.f. Eq. \ref{toomreq}) throughout its evolution even as mass
flows outward.

\subsubsection{Mass Flux}
\label{massflux}

In their study of the similarity solutions of viscously evolving disks,
\cite{lyndenbell74} found that the natural behavior of an accretion disk
is to diffuse and spread out, creating a net inward flow in the inner
disk, a turnaround radius, and an outward decretion flow in the outer
disk.  \cite{ruden86} uncovered the same behavior in numerical
simulations of the evolution of the minimum-mass solar nebula.  Why
haven't many other $\alpha$-disk models predicted the outward mass flow
into the trans-Saturnian region that could so improve the prospects for
giant planet formation?  Most previous evolving disk models such as HGH
and the pseudoevolutionary sequence of \cite{bell97} have required the
disk to be in steady state.  In steady-state models, each annulus
receives an equal inward-driven mass flux with the sun serving as the
sink particle.  There is therefore no chance for any part of the disk
interior to $R_{\rm out}$ at $t = 0$ to gain mass.  Note that the
assumption of steady-state, net inward accretion is still consistent
with viscous expansion of the disk from its original outer boundary, as
occurs in the HGH models and this work.

The non-steady-state evolving disk model of \cite{ruden91} assumed
turbulent stresses were caused solely by convection, and set the
viscosity to zero in optically thin regions where radiative transport is
efficient.  Far from the star, the shallow potential well does not allow
the disk to attain high volume densities, so reaching an optical depth
of $\tau > 1$ requires more mass than in the inner disk.  At $R =
30$~AU, the outer radius at the beginning of the Ruden and Pollack
calculation, convection ceases almost immediately and the disk becomes
quiescent.  The inward accretion flow then removes enough mass from
progressively interior annuli to make them optically thin, leaving
behind a remnant surface density profile where $\Sigma$ {\it increases}
with $R$.

In the Ruden and Pollack simulation, mass is not redistributed from the
inner disk to the outer disk---the slope of $\Sigma(R)$ changes from
negative to positive because the inner parts of the disk remain
optically thick for the longest, participate in convection-driven
accretion the longest, and therefore lose the most mass.  However, in
their analysis of the coupling between the kinetic stress tensor
governing hydrodynamic stability, the magnetic stress tensor, and the
shear field, \cite{hawley99} find that MRI is uniquely capable of
providing turbulent angular momentum transport in disks.  By relaxing
the steady-state condition, but still permitting angular momentum
transport by MRI in optically thin regions (though MRI ceases if ions
and neutrals decouple; see \S \ref{raddiff}), we allow outward flows to
develop and viscous stress to redistribute mass from the inner disk to
the giant planet-forming region.

\begin{figure}
\centering
\includegraphics[scale=0.9]{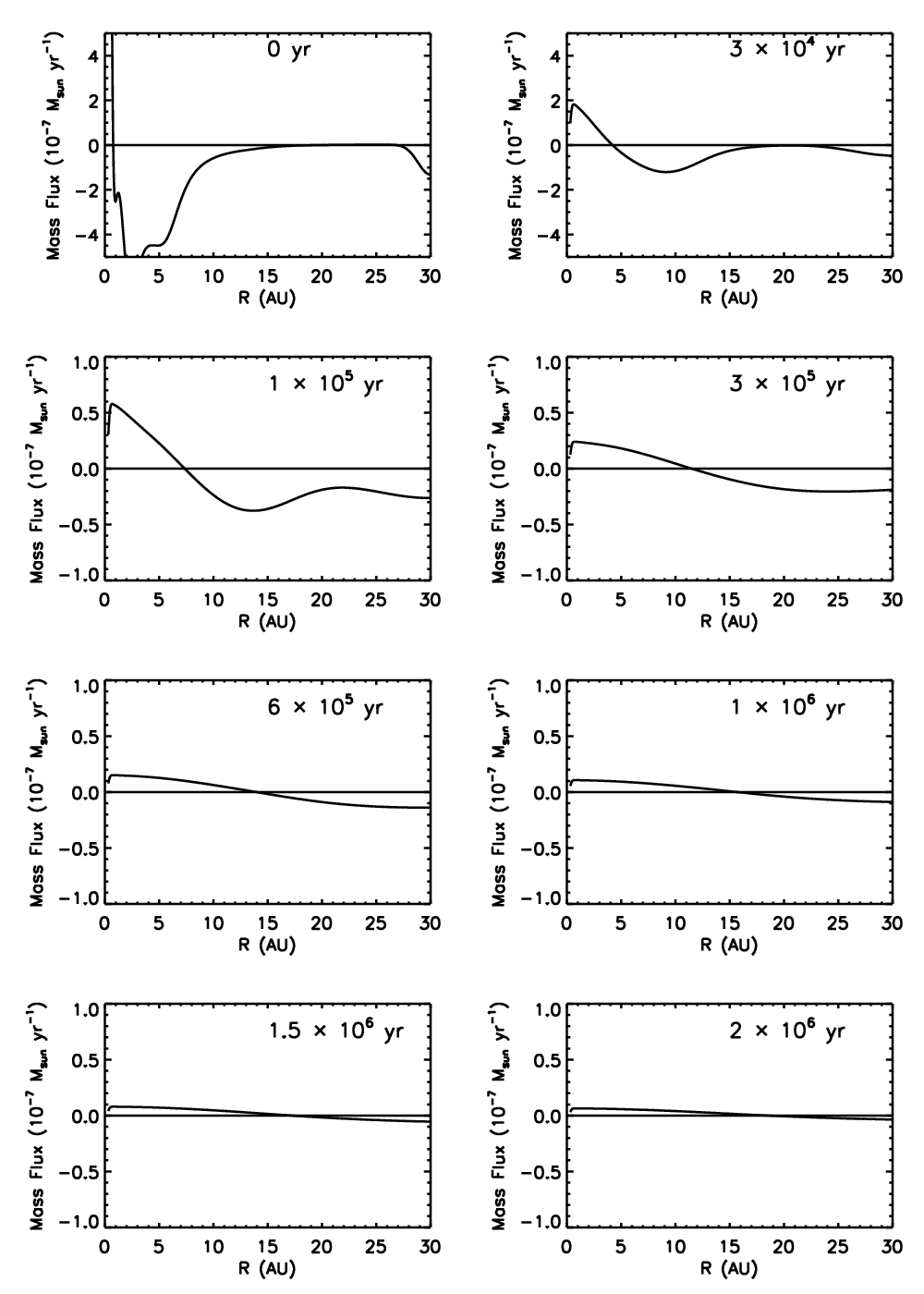}
\caption{Time snapshots of accretion rate as a function of radius.
Negative values of $\dot{M}$ indicate ``decretion'' regions of the disk
with net outward mass flow.}
\label{mdot}
\end{figure}

Figure \ref{mdot} shows time snapshots of the mass accretion rate as a
function of radius.  The location with $\dot{M} = 0$ is the turnaround
between inward accretion and outward ``decretion.'' This turnaround
radius moves outward over time, so that more and more of the disk is
accreting toward the star.  This behavior is required to open a hole in
the inner disk and make the transition from classical T-Tauri to
non-accreting, weak-lined T-Tauri star.  Instead of accreting onto the
star, the outwardly decreting material ($R > 15$~AU) will eventually be
lost to photoevaporation (Alexander et al. 2006).


\subsubsection{Midplane Temperature}

\begin{figure}
\centering
\includegraphics[scale=0.9]{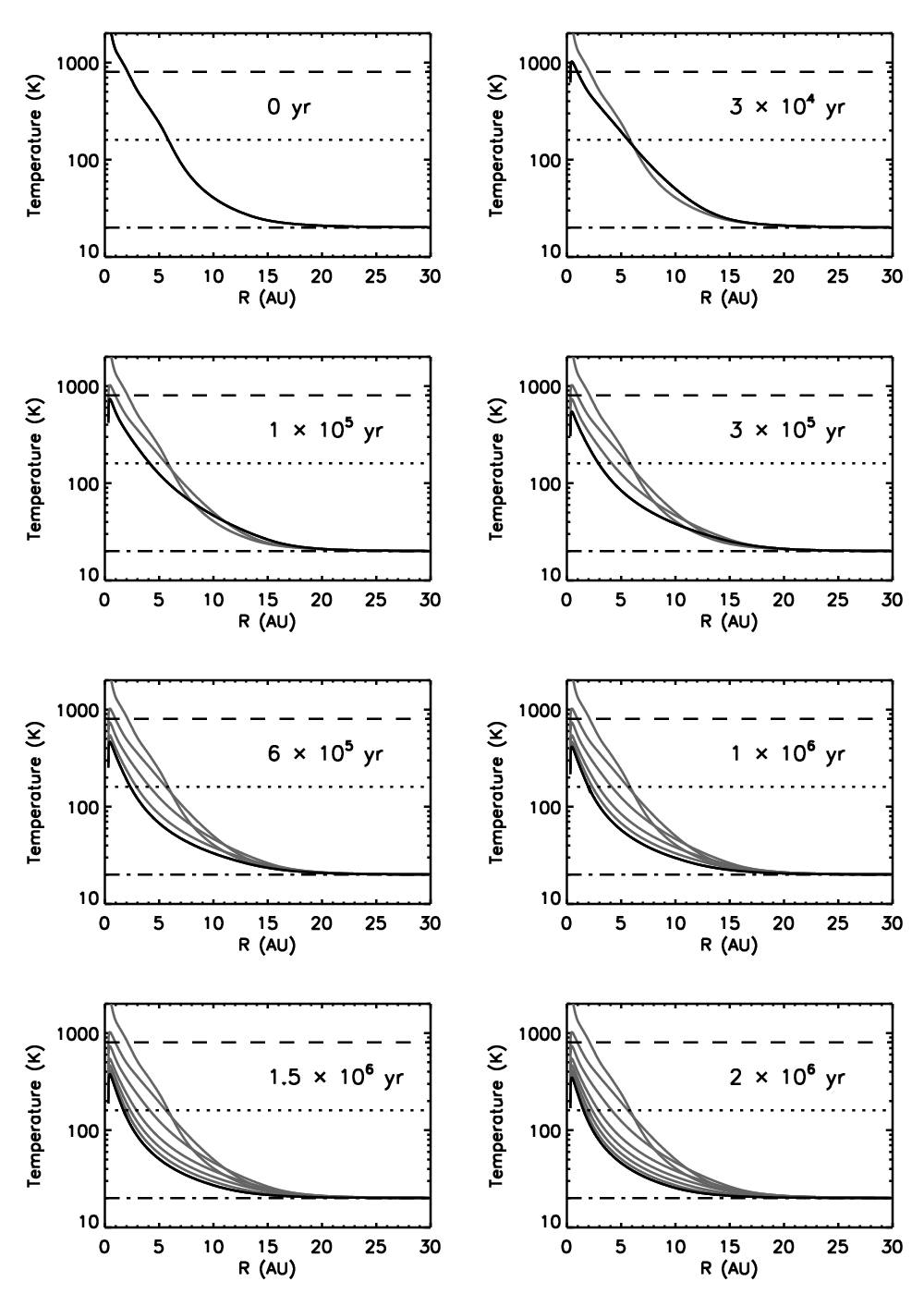}
\caption{Midplane temperature as a function of radius and time in the
inner 30~AU of the disk.  Outside of 30~AU, the midplane is isothermal
at 20~K, the assumed ambient temperature.  The dashed line at 800~K
shows the minimum temperature for silicate crystallization (Gail 1998);
the dotted line at 160~K shows the sublimation temperature of H$_2$O
from the temperature-programmed desorption experiments of
\cite{collings04}; and the dash-dotted line at 20~K shows the assumed
ambient temperature.}
\label{midplanetemp}
\end{figure}

Figure \ref{midplanetemp} shows shapshots of the temperature of the
protostellar disk midplane at $R < 30$~AU as the disk evolves.  Outside
of 30~AU, the entire disk (midplane and above) is isothermal at the
ambient temperature throughout the simulation (20~K, plotted with a
dash-dotted line).  The first three panels of Fig. \ref{midplanetemp}
show that during Epoch 1, areas of the disk that increase in mass are
correspondingly heated: the temperature at 10~AU increases from 40~K at
$t = 0$ to 49~K at $3 \times 10^4$~yr.

We first evaluate the suitability of our model disk for delivering
crystalline silicates to comets, as detected by \cite{keller06} in comet
81P/Wild 2 dust captured by the Stardust spacecraft.  The dashed line in
Fig. \ref{midplanetemp} shows the minimum temperature for silicate
crystallization of 800~K (Gail 1998).  (Note that forsterite formation
requires $T > 1067$~K; Hallenbeck et al. [2000].)  The inner nebula
retains temperatures above 800~K for $10^5$~yr.  \cite{dullemond06}
calculated the formation and transport rates of crystalline silicates in
a 1-d evolving $\alpha$-disk model with infall from a collapsing cloud.
Relaxing the steady-state assumption, they found that viscous spreading
of the disk, rather than turbulent radial mixing, could relocate
crystalline silicates produced early on in the inner, hot ($T \ga
800$~K) parts of the disk to distances of $R > 15$~AU, where comets
formed.  A gradual outward push of crystalline silicates is consistent
with the nebula flow pattern observed in this work, and may explain the
anticorrelation between disk age and crystallinity in the inner disk ($R
< 10$~AU) discussed by \cite{vanboekel04}.  However, the danger that
crystalline silicates transported outward through the midplane would be
swept up by protoplanets before ever reaching the comet-forming region
led \cite{ciesla07} to favor more rapid transport than the decretion
flow in this model.





The dotted line on Fig. \ref{midplanetemp} shows the H$_2$O ice line
at 160~K (Collings et al. 2004).  Although we require a chemical kinetic
model to determine the exact position and width of the ice line (since
ice deposition is not instantaneous but proceeds at a finite rate
limited by the gas velocity of H$_2$O and the available grain surface
area), we can nevertheless estimate its location and compare with
previous disk models.  In our simulation, the H$_2$O ice line is always
inside the formation zones of Saturn, Uranus and Neptune, and crosses
Jupiter's feeding zone (5.2~AU) at $5 \times 10^4$~yr.  All giant
planet cores, therefore, should have formed from icy grains.


After 2~Myr, the midplane outside 15~AU has become isothermal at the
ambient temperature.  This temperature, therefore, has a profound effect
on the ice inventory of the solar nebula.  The most volatile ices (CO,
CH$_4$ and N$_2$) have sublimation temperatures below 45~K.  Simply
raising the ambient temperature above this value would ensure that none
of these ices could freeze on grains.  However, C/H enrichments of
20--40 over the solar value measured in Uranus and Neptune's atmospheres
indicate that direct gas accretion from the nebula cannot explain these
planets' composition (Baines et al. 1995).  We therefore postulate that
the outer solar nebula midplane contained CO and CH$_4$ ice, making our
choice of ambient temperature appropriate.  Further evidence for a cold
solar nebula comes from the observed noble gas enrichment in Jupiter's
atmosphere---trapping argon in icy planetesimals requires temperatures
of 30~K (Owen et al. 1999).


In the next section, we describe our chemical model of the solar nebula
midplane.  We will use these results, along with the surface density
curves shown in Fig. \ref{sdplanet}, to predict the solar nebula's
solid mass distribution in \S \ref{planetesimal}.

\section{Chemical Model}
\label{chemmodel}

\subsection{Reaction Set}

The foundation of our chemical model is the UMIST database RATE95
(Millar et al. 1997).  We follow the chemistry of 211 species, 153
gas-phase and 58 ices.  Our reaction network contains 2479 reactions,
including thermal desorption, gas-grain reactions and grain surface
reactions.  No non-thermal desorption processes (e.g. cosmic ray
heating) are included.

We make the simplifying assumption that radial and vertical motions of
gas and dust are slow compared to chemical reaction timescales, and do
not track the motion of gas and dust through the nebula.  This
assumption is true for grain surface reactions and freezeout/desorption,
which have short timescales, but gas-phase chemistry proceeds more
slowly.  Furthermore, inward migration of cm-size particles can deplete
the outer solar nebula of ice and concentrate each species at its
condensation front (e.g. Ciesla and Cuzzi 2006).  In this paper, we
ignore relative motions of solid and gas.

Since protostellar disks have a high optical depth to UV radiation, we
neglect photoionization in the disk midplane model.  However, we include
cosmic rays at the standard interstellar flux in parts of the disk with
a half-plane surface density lower than the penetration depth of
150~g~cm$^{-2}$ (Umebayashi and Nakano 1981).  The cosmic-ray
penetration depth is listed in Table \ref{freechem} with the other free
parameters in the chemical model.

The following sections describe our treatment of gas-grain and grain
surface reactions and the binding energies we use to calculate those
reaction rates.

\subsubsection{Grain Surface Reactions}

Following \cite{willacy06} and \cite{garrod06}, we use rate equations to
calculate grain surface reaction rates.  Rate equations work in the
mean-field approximation, neglecting the stochastic variation of
abundances on different grains.  The accuracy of rate equations is
compromised when a species reacts on the grain surface faster than it
can adsorb or desorb: this drives the reactant abundance per grain to
less than unity and the mean-field approximation is no longer valid
(Caselli et al. 1998).  However, unlike stochastic methods that account
for the discrete nature of the grains, solving rate equations is
computationally tractable even for large-scale simulations.  For
example, a model that uses direct integration of the master equation
(the probability distribution governing reaction rates) has the number
of equations in the simulation increasing exponentially with the number
of species (Barzel and Biham 2007).

The species most likely to violate the mean-field approximation is
atomic hydrogen, which can scan grain surfaces and find reaction
partners quickly due to its low atomic weight.  While heavier atoms move
across grain surfaces by thermal hopping, Hornekaer et al. (2003) find
that H atoms often tunnel between adjacent vacancies in the grain
surface lattice.  This result conflicts with previous work by Katz et
al. (1999), who found that the quantum tunneling mechanism is
improbable.  Although the question of whether quantum tunneling or
thermal hopping is primarily responsible for hydrogen mobility on grain
surfaces is not resolved, Ruffle and Herbst (2000) show that rate
equation simulations limiting H motion to thermal hopping only---and
thus slowing down reactions involving atomic H---show good agreement
with stochastic methods.  We therefore assume thermal hopping regulates
all grain surface reaction rates.

The thermal hopping rate to an adjacent vacancy on the grain surface
lattice is
\begin{equation}
t_{\rm hop}^{-1} = \nu_0 \exp \left ( -0.3 E_D / k T_g \right ) ,
\label{hop}
\end{equation}
where $E_D$ is the binding energy, $\nu_0$ is the oscillation frequency
between the ice and the grain surface, and $T_g$ is the grain
surface temperature (assumed to be in equilibrium with the gas).  The
oscillation frequency is
\begin{equation}
\nu_0 = \sqrt{2 n_s E_D / \pi^2 m} ,
\label{osc}
\end{equation}
where $n_s \sim 1.5 \times 10^{15} {\rm cm}^{-2}$ is the surface density
of lattice vacancies (Hasegawa et al. 1992; Table \ref{freechem}) and
$m$ is the reactant mass.  Typical oscillation frequencies are
1--3$\times 10^{12}$~s$^{-1}$ (Herbst et al. 2005).  Following
\cite{willacy06}, we assume that only atoms can move across grain
surfaces: molecules, which have higher binding energies, are stationary.
All grain surface reactions thus involve at least one atom.

The grain-surface reaction rate ($s^{-1}$) is given by
\begin{equation}
t_{\rm reac}^{-1} = { t_{\rm hop,1}^{-1} + t_{\rm hop,2}^{-1}
\over 4 \pi a^2 n_s n_g } ,
\label{surfrate}
\end{equation}
where $t_{\rm hop,1}$ and $t_{\rm hop,2}$ are the thermal hopping rates
of reactants 1 and 2, $a$ is the grain radius and $n_g$ is the number
density of grains.  As in the molecular cloud model, we assume a mean
grain radius of $0.258 \; \mu m$ (Table \ref{freechem}).  We take a
grain number density of $n_g / (n_{\rm H} + 2 n_{\rm H2}) = 10^{-12}$
(Table \ref{freechem}), consistent with our assumption of well-mixed
grains and gas.

\subsubsection{Gas-Grain Reactions}

We assume that all species can freeze out on grains except He, which has
a very low binding energy, $E_D / k = 100$~K (Tielens and Hagen 1982).
The freeze-out rate is
\begin{equation}
k_f = S_x \left \langle \pi a^2 n_g \right \rangle C_i v_x n_x ,
\label{freezerate}
\end{equation}
where $S_x$ is the sticking coefficient (assumed to be 0.3 for H atoms
and 1 for all other species; Table \ref{freechem}), $v_x$ is the
gas-phase velocity of species $x$, and $n_x$ is the gas-phase number
density of $x$.  The factor $C_i$ takes into account the attraction of
positively charged ions to grains which, on average, are negatively
charged (Umebayashi and Nakano 1980):
\begin{equation}
\left \{ \begin{array}{l} C_i = 1 \; {\rm for \; neutrals} \\ C_i = 1 +
e/(akT) \; {\rm for \; single-charged \; positive \; ions.} \end{array}
\right.
\label{ci}
\end{equation}
In Eq. \ref{ci}, $k$ is the Boltzmann constant and $e$ is the
electron charge.  Once an ion hits a grain, we assume it is neutralized
by an adsorbed free electron.  The thermal desorption rate ($s^{-1}$) at
which ice mantle species are returned to the gas is
\begin{equation}
k_t = \nu_0 \exp \left ( E_D / k T_g \right ) .
\label{therm}
\end{equation}


\subsubsection{Binding Energies}

The binding energies $E_D$ for each species depend on the surface onto
which the molecules adsorb and the mixture of ices on that surface.  If
each ice were a pure solid, $E_D$ would simply be the sublimation energy
and desorption physics could be calculated nonkinetically using vapor
pressure.  However, molecules in mixed ices may have substantially
different binding energies than those in pure ices, so the sublimation
energy approximation is inadequate.

The most abundant astrophysical ice is H$_2$O, so the binding energy of
each species under protostellar disk conditions is primarily determined
by how it behaves on/in an amorphous water-ice matrix.
\cite{collings04} group binding energies into three categories based on
their behavior when co-deposited with H$_2$O: CO-like, water-like, and
intermediate.  The extremely volatile CO-like species include CO, N$_2$,
O$_2$, and CH$_4$.  Molecules of these species do not bond efficiently
either with themselves or with water: they quickly diffuse through an
amorphous H$_2$O matrix and desorb mainly at their sublimation
temperatures, which are quite low ($<40$~K), with minor amounts
remaining trapped in the H$_2$O ice.

Molecules of water-like species, including NH$_3$ and CH$_3$OH, bind
more strongly to water than to each other or to a metal substrate.
These polar molecules form hydrogen bonds with water that keep them in
the solid phase at higher temperatures than possible in their pure form.
The data of \cite{collings04} show that ammonia-water mixtures are
substantially less volatile than pure ammonia, which has a binding
energy of only 1110~K (Hasegawa and Herbst 1993).  NH$_3$ therefore takes
on the binding energy of H$_2$O, 5700~K.

Finally, intermediate species are trapped by H$_2$O to some extent, but
do not have the same mobility as CO inside a water-ice matrix.  Some
co-desorption with H$_2$O occurs, but most desorption takes place at the
sublimation energy.  These species include H$_2$S, OCS and CH$_3$CN.  We
have not included the partial co-desorption with water and treat each
molecule as having only one specific desorption tmeperature.  Table
\ref{binding} gives the binding energy data used in our model.

\subsection{Preprocessing in Molecular Cloud}

The ices in protostellar disks have undergone significant chemical
processing before giant planet formation begins.  To simulate this
preprocessing, we model $10^6$ years of molecular cloud evolution to
derive input abundances for the protostellar disk.  The molecular cloud
model uses the same species set and reactions as our disk model, with
the addition of photoionization due to the interstellar UV field.

The atoms that form the species in our simulation are H, He, C, N, O,
and S.  We also include Si adsorption, desorption and ionization as a
tracer of the metal population.  Except for hydrogen, we assume all
material enters the molecular cloud in atomic form, as if coming from a
recent supernova.  We take $n_{\rm H} / (n_{\rm H} + 2 n_{\rm H2}) =
10^{-2}$ (Table \ref{freechem}), so that 99\% of hydrogen atoms are in
H$_2$.  We assume a density of $2 \times 10^4$~cm$^{-3}$, a temperature
of 10~K and an extinction $A_V = 10$~mag.

Not all of the atomic CHONS material is available to form ices: some
will help form the solid grains where ice mantles deposit.  To figure
out the fraction of C, N, O, and S locked away in grains, we use the
atomic abundance ratios in Halley's comet dust grains reported by
\cite{jessberger88}: C:N:O:Si:S = 814:42:890:185:72.  Assuming only
trace amounts of silicon remain in the gas phase (gas/grain $< \:
0.001$), we calculate grain fractions for C, N, O and S as
\begin{equation}
n_{\rm grain} / n_{\rm gas} = \left ( {N_{\rm Si} \over N_x} \right
)_{\odot} \left ( {N_x \over N_{\rm Si}} \right )_{\rm comet} .
\label{dustfrac}
\end{equation}

Using the solar composition of \cite{helling00}, as in the dynamical
disk model, we find $F_{\rm grain}$ = 0.4, 0.1, 0.2 and 0.85 for C, N, O
and S respectively (Table \ref{freechem}).  This portion of each atom's
inventory does not participate in chemistry and is excluded from the
simulation.  The gas and ice abundances of major species after 1~Myr of
molecular cloud evolution are listed in Table \ref{cloudoutput}.  These
abundances form the chemical initial conditions for our protostellar
disk model.

\begin{figure}
\centering
\begin{tabular}{c}
\includegraphics[scale=0.5]{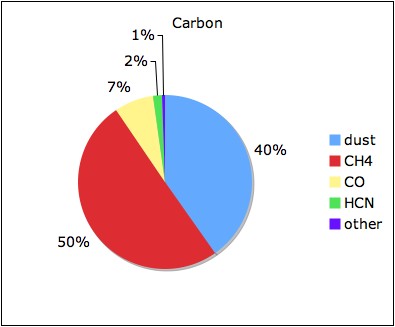} \\
\includegraphics[scale=0.5]{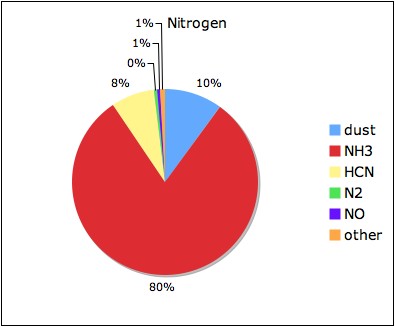} \\
\includegraphics[scale=0.5]{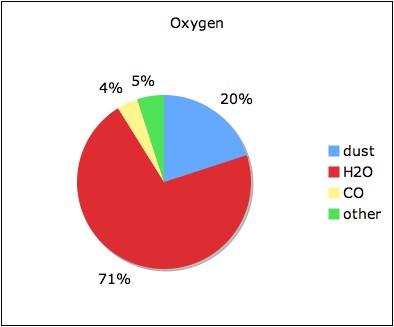} \\
\end{tabular}
\caption{Distribution of C, N and O atoms among dust and most common ice
species.  Proportions are rounded to the nearest 1\%.}
\label{pies}
\end{figure}

The major ice species formed in the molecular cloud are H$_2$O, NH$_3$,
CO, and CH$_4$.  More than 90\% of the C, N and O atoms that aren't part
of refractory grains are in these compounds.  Sulfur is sequestered in
solid H$_2$S after the molecular cloud phase, but this composition is
subsequently modified during disk evolution.  Minor CNO carriers are
HCN, N$_2$ and NO.  Figure \ref{pies} shows the division of solid C, N
and O atoms among the species in our model.

Our molecular cloud model predicts total ${\rm CO / CH_4} = 0.14$.  Both
of these species reside in ice mantles.  Abundances of major ices
relative to H$_2$O are listed in Table \ref{iceratios}.  Our CO/CH$_4$
ice ratio is consistent with the \cite{aikawa08} models of prestellar
cores, which give ${\rm CO / CH_4} = 0.43$ in an embedded core and ${\rm
CO / CH_4} = 0.1$ in an isolated core.  Likewise, Lodders (2003)
concluded that the reactions
\begin{equation}
\begin{array}{ll}
3 {\rm H_2 + CO} \rightarrow {\rm CH_4 + H_2O} \\
{\rm CO} \rightarrow {\rm C} \; ({\rm graphite}) + {\rm CO_2}
\end{array}
\label{noco}
\end{equation}
would deplete CO from presolar gas (note, however, that in our model
CH$_4$ forms on grains and not in the gas).  Lodders predicts ${\rm CH_4
/ H_2O} = 0.58$, whereas we find a methane-to-water ratio of 0.38 (Table
\ref{iceratios}).  This discrepancy reflects the fact that we have a
higher input oxygen abundance (to match the opacity tables calculated by
Semenov [2003]).

We note that the dominant components of interstellar ice mantles are
typically H$_2$O, NH$_3$, CO, CO$_2$, and CH$_3$OH (Charnley and Rodgers
2008), which suggests carbon oxide formation in the ISM should be
favored and hydrocarbon formation kinetically inhibited.  One possible
way to reduce the ${\rm CH_4 / CO}$ ratio is to lower the assumed atomic
hydrogen abundance, which would reduce the efficiency of hydrogenated
molecule formation on the grains at early times when atomic carbon is
abundant.  However, the purpose of this work is to determine the ice
mass available for giant planet formation. Since CH$_4$ and CO have such
similar condensation temperatures (41 and 35 K, respectively) and we
assume ${\rm O/C} > 1$, so CO cannot incorporate all available oxygen,
${\rm CO/CH_4}$ is not critical to our final result.

We find that the most abundant nitrogen-carrying molecule is ammonia,
with ${\rm N_2 / NH_3} = 2.5 \times 10^{-3}$ (both species reside in ice
mantles).  We calculate ${\rm NH_3 / H_2O} = 0.14$, which matches the
input N/O ratio.
Lodders (2003) also predicted that hydrated ammonia was the major
nitrogen reservoir in the solar nebula (see \S \ref{ammonia} for a
discussion of hydrated ammonia in Jupiter and Saturn's feeding zones).
However, \cite{womack92} infer ${\rm N_2 / NH_3} = 4$ in the
comet-forming region of the solar nebula from models of elemental
nitrogen depletion in Comet Halley.

Overall, our model molecular cloud favors the formation of hydrogenated
C, N, and O compounds over oxides or diatomic molecules (except for
H$_2$).  We next discuss the subsequent chemical evolution of the solar
nebula.

\subsection{Chemical Model Results: Ice Lines} 
\label{icelinesec}

In this section, we report the locations of important condensation
fronts and the relative abundances of the most common ices.  We use the
gas and ice mixture resulting from the molecular cloud model as the
starting point for a radial series of midplane chemical models that span
2~Myr of disk evolution.  These models track the ice inventory of
the disk midplane as a function of heliocentric distance and time.

The ices in the disk can be subdivided into two categories: (1) those
that form in the molecular cloud and freeze out from the disk gas, and
(2) those that form or are destroyed in the warm disk midplane.  All
Category 1 molecules are highly stable both as gases and solids: in the
radiation-shielded disk midplane, their reaction probabilities are very
low.  The stability and ease of formation of these molecules makes them
highly abundant, $> 99$\% of the solar nebula ice mass.  In Category 1
are water, methane, CO, ammonia, and HCN. In Category 2 are all
sulfur-carrying molecules, simple hydrocarbons, N$_2$, and NO.

\subsubsection{Category 1: Stable and Abundant}

\begin{figure}
\centering
\includegraphics[scale=0.9]{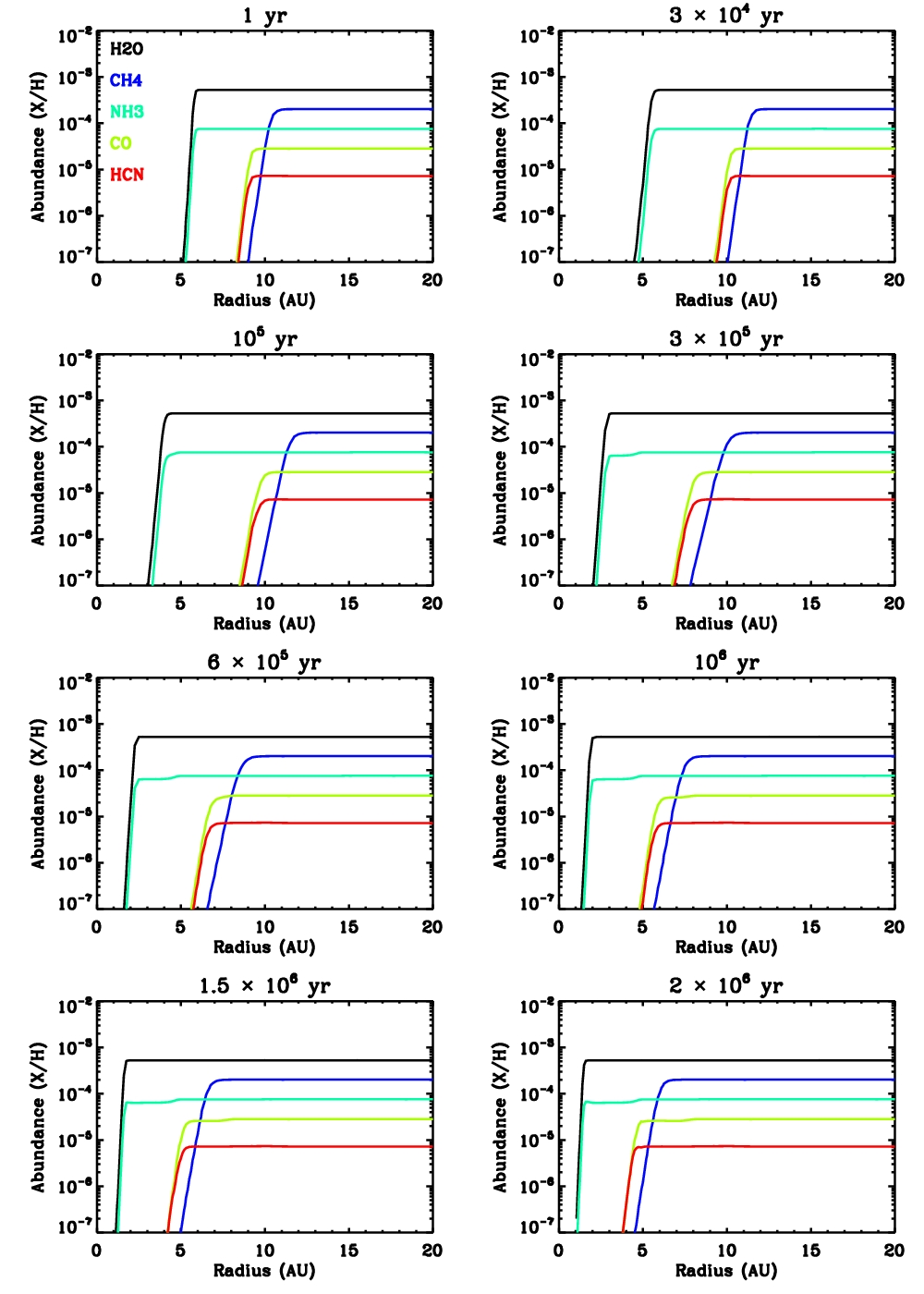}
\caption{Time snapshots of ice abundance as a function of radius for
species formed mainly by gas freezeout.  Note that the condensation
fronts are not step functions: the region of partial ice freezeout can
cover up to 3 AU.}
\label{freezeabun}
\end{figure}

Figure \ref{freezeabun} shows time snapshots of the abundance of
Category 1 ice abundances (solid phase only) as a function of radius.
Note that the ammonia and water condensation fronts coincide; we will
discuss the implications of this further in \S \ref{ammonia}.  The most
abundant ice is H$_2$O, followed by methane, ammonia, CO and HCN.  These
five ices account for 98\% of the ice mass in regions where $T \leq
35$~K, the sublimation temperature of CO.

In Fig. \ref{freezeabun}, we see the H$_2$O condensation front sweeping
from 6~AU at $t = 0$ to 1.5~AU after 2~Myr.  Observations of the
asteroid belt place the H$_2$O ice line at 2.7~AU (Abe et al. 2000;
Rivkin et al. 2002), a position it reaches at $t = 5 \times 10^5$~yr in
our model.  The slow motion of the snow line through the inner solar
system could perhaps account for the H$_2$O abundance gradient observed
in the asteroid belt (Barucci et al. 1996): we find that icy grains are
present for $\sim 10^5$ years longer at 3~AU than at 2~AU.  The snow
line crosses Jupiter's heliocentric distance, 5.2~AU, after only $5
\times 10^4$ yr, leaving ample time for the formation of Jupiter's core
from icy planetesimals.

If we assume Saturn formed {\it in situ} at 9.5~AU, primordial methane
accretion onto both the giant planet and its satellites is possible.  We
find that grain mantles in Saturn's feeding zone should have ${\rm CH_4
/ H_2O} \ge 0.01$ after $2.3 \times 10^5$~yr.  CH$_4$ is 90\% solid at
9.5~AU after $5.4 \times 10^5$~yr.  We note that the C/H ratio in
Saturn's atmosphere has recently been revised upward by a factor of 2
(Flasar et al. 2005).

We find that if Uranus and Neptune formed outside 12~AU, their
core-forming planetesimals contained both CH$_4$ and CO.  The
composition of both planets strongly suggests that they accreted
carbonaceous ices: ${\rm (C/H) / (C/H)}_{\odot}$ is 41 for Neptune and
$\leq 260$ for Uranus (Lodders and Fegley 1994).  Furthermore, Grundy et
al.  (2002) find that methane ice is widely distributed across the
surface of Triton.

\subsubsection{Category 2: Chemically Active}

In Fig. \ref{freezeabun} we see that the typical behavior of an ice
line is to move inward as the nebula cools.  This can happen only if the
molecule is relatively inert.  Sulfur species are missing from Fig.
\ref{freezeabun} because H$_2$S, the dominant sulfuric ice at $t = 0$
with 99.6\% of the sulfur atoms, does not survive in the gas phase.
Likewise, while C$_3$H$_n$ chains are stable and freezeout-dominated in
our model, acetylene ice partially hydrogenates to form ethylene and
ethane.  Gaseous NO and N$_2$ participate in ion exchange reactions in
the gas, reducing the abundance available for ice formation.  In this
section, we consider the sulfuric, nitric and hydrocarbon ice systems.

\begin{figure}
\centering
\includegraphics[scale=0.6]{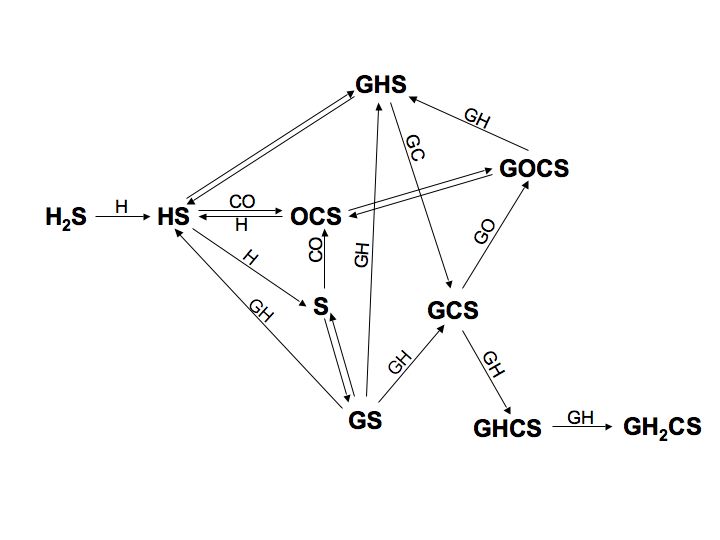}
\caption{Reactions linking the most abundant sulfuric ices, H$_2$S,
H$_2$CS and OCS.  The label ``G'' in front of a compound name denotes
that the species is solid, adsorbed on the grains.  The notation $H_2S
\stackrel{H}{\rightarrow} HS$ denotes the reaction $H_2S + H \rightarrow
HS + H_2$; all other reaction pathways can be read analogously to this
example.  Non-sulfuric reaction by-products, such as H$_2$ in the
aforementioned reaction, are not shown.  An unlabeled double arrow, such
as the one connecting GOCS and OCS, denotes a phase change.  Reaction
pairs are not in equilibrium.}
\label{sulfurnetwork}
\end{figure}

\begin{figure}
\centering
\includegraphics[scale=0.9]{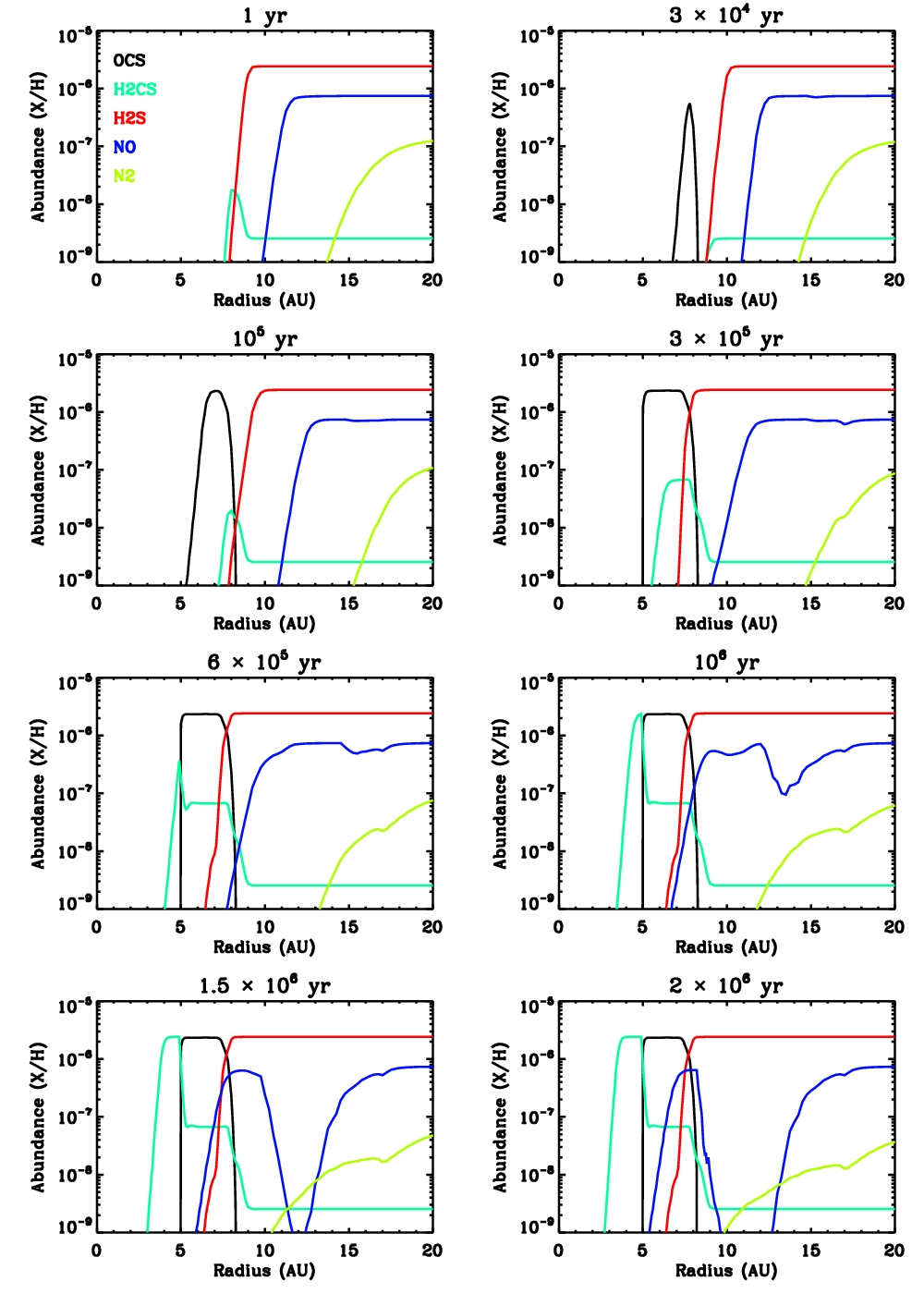}
\caption{Abundances of solid NO, N$_2$, and sulfur-carrying species as a
function of radius and time.  We find a banded sulfur ice distribution
in which H$_2$S dominates outside $\sim 7$~AU, OCS is most abundant
between 5 and 7~AU, and H$_2$CS is dominant in the inner nebula.}
\label{sulfuric}
\end{figure}

As soon as our midplane chemical model begins, the reaction
\begin{equation}
{\rm H_2S + H \rightarrow HS + H_2}
\label{h2s}
\end{equation}
starts destroying gaseous H$_2$S where temperature is above $\sim 50$~K.
The product HS then initiates a network of reactions that form OCS and
H$_2$CS.  The reactions linking the main sulfur-carrying molecules are
shown in Fig. \ref{sulfurnetwork}.  We do not find any depletion of
H$_2$CS once formed on grains.  Figure \ref{sulfuric} shows time
snapshots of the abundances (solid phase only) of the main
sulfur-carrying molecules H$_2$S, OCS and H$_2$CS.  After 2~Myr, we see
a banded structure where H$_2$S dominates at $R > 8$~AU, OCS is most
abundant between 5 and 8~AU and H$_2$CS dominates at $R < 5$~AU.

H$_2$S remains in the outer solar nebula because the gas at $R > 9$~AU
never gets hotter than the H$_2$S sublimation temperature in our model.
This scenario may be physically unrealistic: infalling molecular cloud
material is likely heated in an accretion shock during disk formation.
However, if we examine the total abundance of icy sulfur atoms, $N({\rm
H_2S}) + N({\rm OCS}) + N({\rm H_2CS})$, we see that it is conserved
after OCS formation finishes at $3 \times 10^5$ yr.  From Fig.
\ref{sulfuric}, we can imagine a sulfurous ice line sweeping through the
solar nebula, moving from 8.5~AU at $t = 0$ to 2.5~AU after 2~Myr.

Figure \ref{sulfuric} also shows the abundances of solid NO and N$_2$.
NO is chemically active during the late stages of the solar nebula, $T
\geq 10^6$~yr.  Even when the solar nebula temperature drops below the
nominal sublimation temperature of NO, some thermal desorption still
takes place, albeit at a much-reduced rate (see Eq. \ref{therm}).
For most species, re-freezing after desorption occurs faster than
reactions in the gas and no ice molecules are lost.  However, NO can be
destroyed in the gas by reactions with ${\rm H^+}$.  Once the half-plane
surface density drops below the cosmic ray penetration depth at late
times, the production rate of ${\rm H^+}$ increases by 33\% due to the
reaction of ${\rm H_2}$ with cosmic ray-produced photons:
\begin{equation}
{\rm H_2} + \gamma_{CR} \rightarrow {\rm H^+ + H + e^-} .
\label{makehplus}
\end{equation}
Thermally desorbed NO atoms are then efficiently removed from the gas
before refreezing by the reaction sequence
\begin{equation}
\begin{array}{c}
{\rm NO + H^+ \rightarrow NO^+ + H} \\
{\rm then} \\
{\rm NO^+} \stackrel{\rm freeze}{\longrightarrow} {\rm N + O} \\
{\rm or} \\
{\rm NO^+ + e^- \rightarrow N + O} . \\
\end{array}
\end{equation}

Laboratory work by \"{O}berg et al. (2005) indicates that N$_2$ is
extremely volatile, with $E_D / k = 790$~K.  This puts its 
sublimation temperature near the assumed ambient temperature of our
protostellar disk simulation, 20~K.  The smooth shape of the N$_2$
abundance curve at $t = 1$~yr reflects the midplane temperature's
asymptotic approach to isothermality at 20~K.  As the disk evolves,
the N$_2$ ice line moves inward, but the overall N$_2$ abundance
declines due to the gas-phase reaction
\begin{equation}
{\rm He^+ + N_2 \rightarrow N^+ + N + He} .
\label{dinitrogen}
\end{equation}
The ${\rm N^+}$ ion then begins a sequence of reactions with H$_2$ that
produce ${\rm NH^+}$, ${\rm N_2H^+}$, ${\rm N_3H^+}$ and finally the
ammonium ion ${\rm N_4H^+}$.  Ammonium recombines with an electron to
form NH$_3$ + H, so the net effect is to convert gaseous N$_2$ into
NH$_3$.

Our model includes the small aliphatic hydrocarbons C$_2$H$_2$,
C$_2$H$_4$, C$_2$H$_6$, C$_3$H$_2$ and C$_3$H$_4$.  We do not include
longer chains in our model---we merely use these five species to find
the location of the ``oil line'' of icy hydrocarbons.  Lodders (2004)
suggested that tarry compounds, rather than water, provided the bulk of
the solid mass for Jupiter's core.  This work does not truly test that
hypothesis, since Lodders' scenario requires a substantial amount of
hydrocarbons that are more refractory than water (our light oils have
sublimation temperatures of 55-70~K).  PAHs would be suitable tar
candidates, but we do not include them in our model for lack of accurate
binding energy measurements.  We plan on incorporating PAH freezeout
into future work, but we note that Jupiter falls inside the water-ice
line for all but the first 50,000~yr of disk evolution.

\begin{figure}
\centering
\includegraphics[scale=0.9]{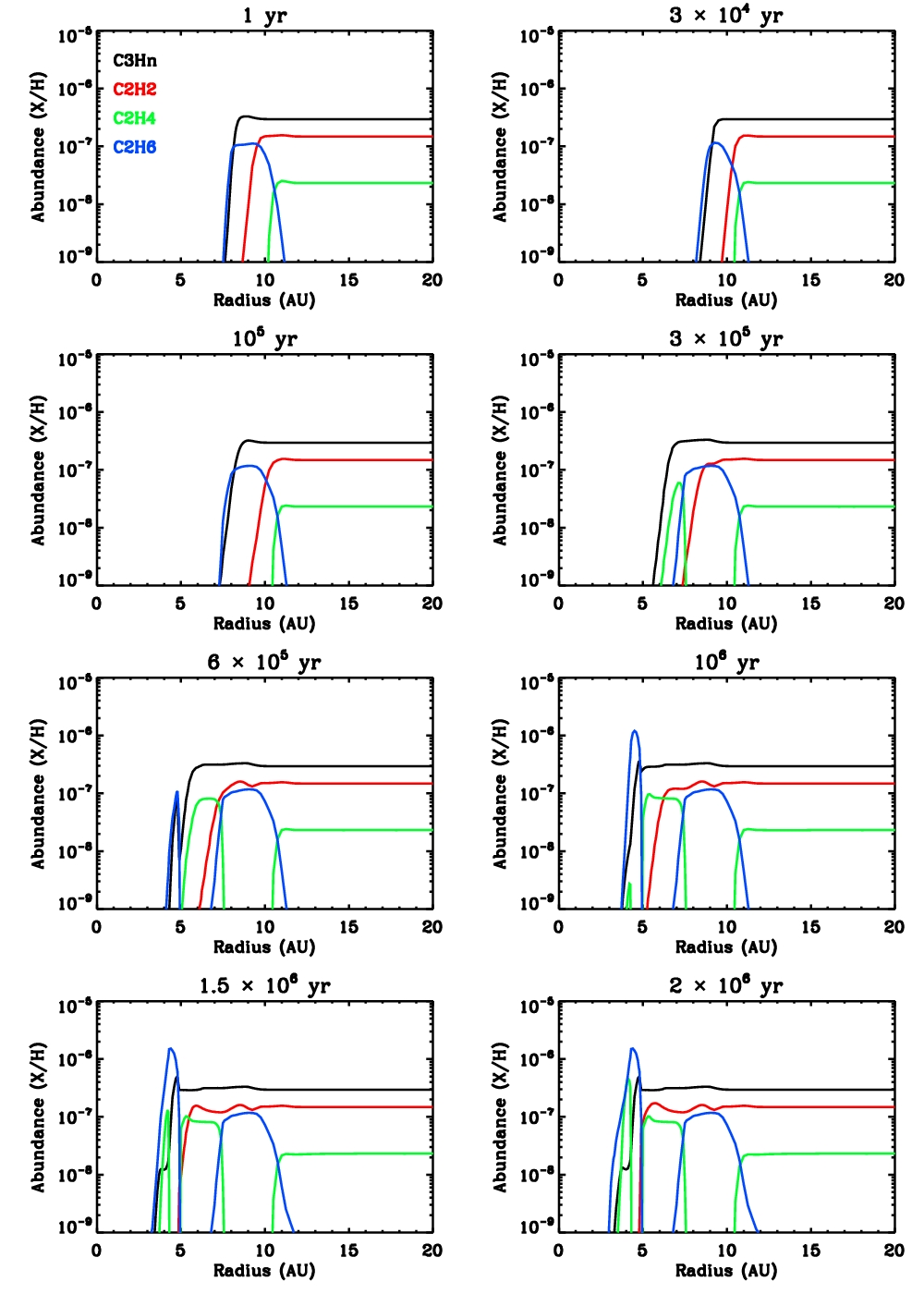}
\caption{Hydrocarbon ice abundances as a function of radius and time.
We group all C$_3$H$_n$ compounds together because our reaction sequence
ends at C$_3$H$_4$, so we cannot differentiate between different levels
of C$_3$ saturation.}
\label{hydrocarbon}
\end{figure}

Figure \ref{hydrocarbon} shows the evolution of the hydrocarbon ice
lines.  The most abundant hydrocarbons in our simulation are C$_3$H$_n$
chains, which are at the end of our reaction sequence.  This suggests
that concatenation of small carbon chains is efficient.  The small
hydrocarbons in our model have similar sublimation temperatures,
creating an ``oil line'' that moves from 8~AU at $t = 0$ to 3~AU after
2~Myr.

For the C$_2$H$_n$ hydrocarbons, acetylene dominates over ethylene and
ethane in the trans-Saturnian region, $R > 10$~AU.  The activation
barrier $E/k = 1210$~K for the reaction ${\rm C_2H_2 + H \rightarrow
C_2H_3}$ (Hasegawa et al. 1992), which has to break the strong C-C
triple bond, means that at the cold temperatures required for acetylene
to deposit on grain surfaces, hydrogenation proceeds slowly.  Only when
the disk is near the acetylene sublimation temperature ($\sim 55$~K)
does a significant amount of ethane form.  Throughout most of the disk,
ethylene is more abundant than ethane.  However, an ``ethylene gap''
forms between 7 and 10~AU after $3 \times 10^4$~yr.  We explain the
ethylene gap as follows: between $10^4$ and $10^5$~yr, 7--10~AU region
takes on mass due to an outward-moving density wave (recall the
discussion in \S \ref{surfacedensity}).  This surface density increase
causes a temperature increase of $\sim 15$~K, increasing the rate of
ethylene hydrogenation.



\subsubsection{The Snow-Ammonia Line}
\label{ammonia}

Since we have adapted lab results showing co-desorption of water and
ammonia (Collings et al. 2004) for our model, we find that the ammonia
and water condensation fronts coincide.  This statement may seem
controversial, and indeed, further lab experiments are needed to verify
that hydrogen bonding between the species is strong enough to govern
ammonia desorption.  However, the composition of Jupiter, Saturn and
their satellites provides strong evidence for primordial ammonia-rich
planetesimals between 5 and 10 AU.

Lopes et al. (2007) examined the rheological properties of a
cryovolcanic flow on Titan with Synthetic Aperture Radar imaging and
found a slurry composition most consistent with an
ammonia-water-methanol mixture (recall that methanol is another ice
found to co-desorb with water in lab experiments).  The presence of
ammonia in cryovolcanic outflows strongly suggests primordial ammonia
accretion: late-stage volatile delivery by comets would provide a small
surface reservoir of ammonia but not embed it in the geologically active
subsurface.  Based on the low temperatures ($T < 75$~K) necessary for
trapping N$_2$ in water ice, \cite{owen00} argued that Titan's massive
nitrogen atmosphere originally supplied as NH$_3$, {\it unmodified from
its relative abundance in the outer solar nebula.}


Other Saturnian moons also show evidence of primordial ammonia
accretion: Prentice (2007) calculates that Iapetus is 27\% ammonia by
mass, assuming homologous contraction of the Saturnian subnebula and
placing Enceladus just at the stability point of liquid water.  The
source material for the rock/ice Iapetus cannot be solar nebula gas, nor
could N$_2$ have provided the initial nitrogen budget: with 34\% rock
and 34\% H$_2$O, there would have been no free hydrogen reservoir for
ammonia formation.  Freeman et al. (2007) suggest that an ammonia-water
ice mantle is necessary to explain the subsurface ocean driving
Enceladus' south pole geyser.  Finally, Prentice (2006) finds that
the lack of compressional features on the surface of Rhea, which are
expected to result from phase II of H$_2$O crystalline ice, favors an
ammonia-rich satellite composition (25\% NH$_3$ by mass), since ammonia
inhibits phase II H$_2$O ice formation.

In the Jovian system, conducting subsurface oceans are required to
explain the magnetic fields of Europa, Ganymede, and Callisto.  One way
to lower water's melting point and maintain liquid oceans is to mix in
ammonia (Hussmann et al. 2006).  Spohn and Schubert (2003) use
equilibrium heat transfer models to confirm the likelihood of subsurface
ammonia-water oceans on all Galilean satellites except Io.

Perhaps the best piece of evidence for ammonia-containing planetesimals
between 5 and 10 AU is the nitrogen enrichment in Jupiter and Saturn's
atmospheres.  ${\rm (N/H) / (N/H)}_{\odot} = 3.3$ for Jupiter and 2--4
for Saturn (Owen and Encrenaz 2003).  Such a large enrichment suggests a
non-gaseous origin for nitrogen in both planets.  Trapping N$_2$ or
NH$_3$ in clathrate hydrates is possible, but \cite{hersant04} find that
the ammonia clathrate cannot form until 0.9~Myr after solar nebula
formation: the ammonia clathrate is more volatile than a co-deposited
ammonia-water ice mixture.  Based on the upward revision of Saturn's C/H
ratio, Hersant et al. (2008) suggest that ammonia was present as a
hydrate, rather than a clathrate, in Saturn's feeding zone.  {\it We
recommend that the canonical ``snow line'' be reconceived as a water
ice-ammonia line, and that NH$_3$ be added to the volatile inventory of
Jupiter- and Saturn-forming planetesimals.}


\section{Solid Surface Density}
\label{planetesimal}

Knowing the solid-phase abundance of each ice species and the location
of its condensation front, we can now calculate the solid surface
density available for planet formation everywhere in the disk.  We first
calculate the gas/solid mass ratio
\begin{equation}
G / S = {\sum_g \mu_g n_g \over \sum_s \mu_s n_s} ,
\label{gsr}
\end{equation}
where $\mu$ is the mean molecular weight of a species, $n$ is its
abundance, and the subscripts $g$ and $s$ denote gas and solid species,
respectively.  We include Na, Mg, Si, Fe, Ni, Al and Ca in the solid
inventory in solar proportions and assume these atoms form refractory
compounds.  Finally, we calculate the solid surface density as
\begin{equation}
\Sigma_{\rm solid} = {\Sigma_{\rm total} \over 1 + G / S} .
\label{sigmasolid}
\end{equation}

\begin{figure}
\centering
\includegraphics[scale=0.9]{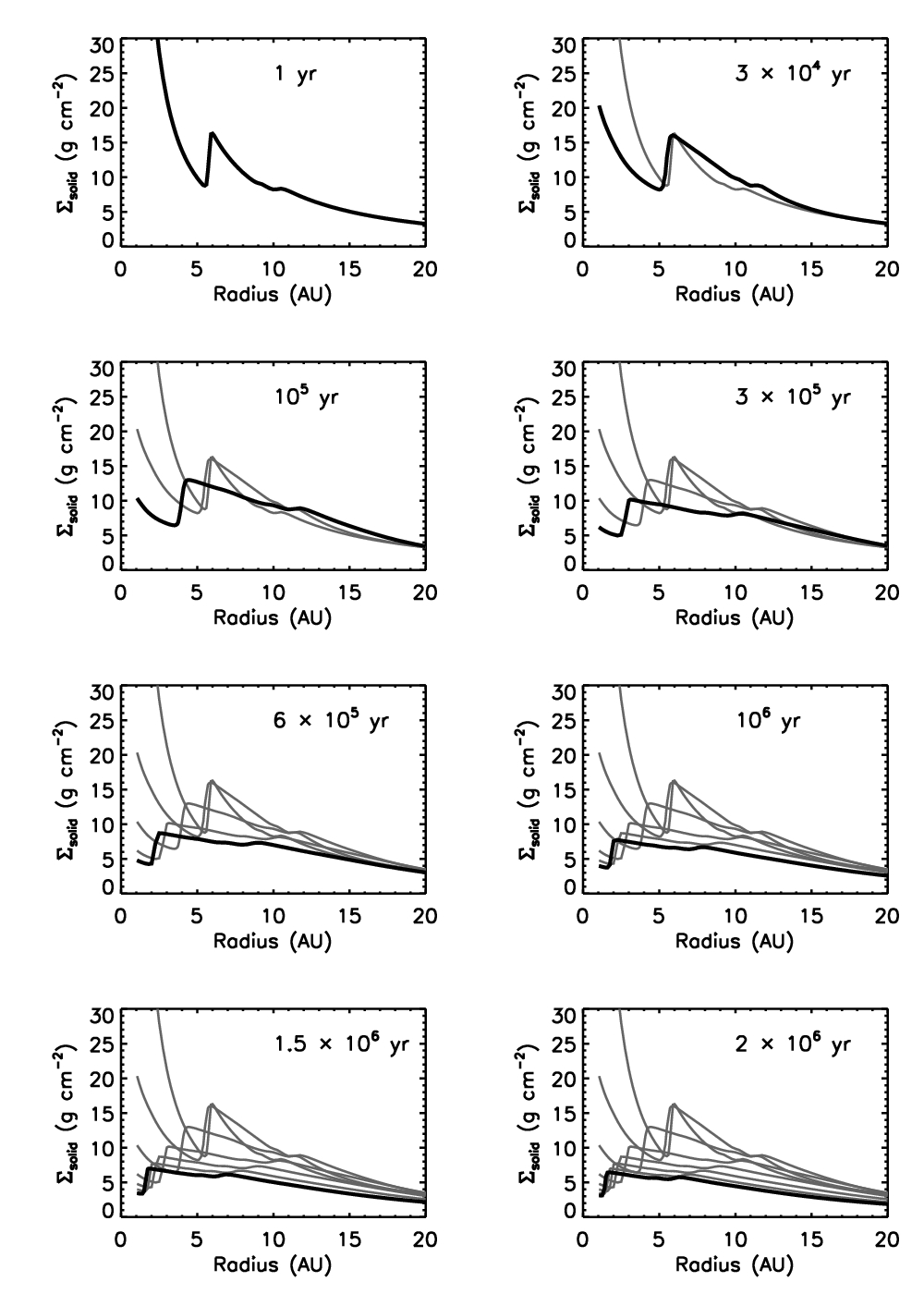}
\caption{Time snapshots of solid surface density available for planet
formation as a function of radius.  In each panel, the current solid
surface density is plotted in black and previous time snapshots are
retained in gray.}
\label{solids}
\end{figure}

Figure \ref{solids} shows the evolution of the solid surface density
distribution in the solar nebula.  Note that as the disk evolves, the
solids should grow and decouple from the gas, whereas these solid
surface density calculations assume grains and gas are well mixed.  We
see the large bump caused by the snow-ammonia line beginning at 6~AU and
moving inward to 1.5~AU after 2~Myr.  In addition, methane adsorption
creates a second local maximum in the solid surface density profile at
11~AU, which moves inward to 6.5~AU by the end of the simulation.

Panels 1 and 2 in Fig. \ref{solids}, $t = 1$ and $3 \times 10^4$~yr,
show the effect of the decretion flow on the solid surface density
profile.  The solid mass between 6 and 14~AU actually increases during
the early stages of disk evolution, at the expense of the inner 4~AU.
The next panel, $10^5$~yr, shows the further solid surface density
increase between 12 and 18~AU.  Disk annuli that increase in surface
density also heat up, and the methane bump consequently moves outward
for the first $10^5$~yr.

\cite{pollack96} presented core accretion models of Jupiter, Saturn, and
Uranus, beginning with solid surface densities of 10, 3, and
0.75~g~cm$^{-2}$.  While these simulations showed the plausibility of
giant planet formation by core accretion, the planet formation
timescales calculated were too long: Jupiter, Saturn and Uranus took 8,
9.6, and 16~Myr to form, respectively.  Since runaway solid core growth
rate is directly proportional to surface density (Eq.
\ref{planetgrowth}), we see that having a high surface density speeds up
giant planet formation.

If one assumes that nearly 100\% of the available solids were
incorporated into planetesimals, our results give starting solid surface
densities that are far more conducive to giant planet formation than
those from \cite{pollack96}.  (The $\Sigma_{\rm solid}$ of
\cite{pollack96} refers specifically to 100-km planetesimals, not
smaller bodies or grains.)  Table \ref{sdtable} lists solid surface
density as a function of radius and time in the solar nebula.  Note that
these results assume a substantial population of small grains is always
available for ice deposition.

We consider $1.5 \times 10^5$~yr the extreme upper limit of the
planetesimal formation timescale, which Hubbard and Blackman [2006]
predict to be $\le 5 \times 10^4$~yr.  However, we continue the disk
simulation out to 2~Myr in order to calculate nebular gas temperature,
density and composition, which are necessary for models of giant planet
formation that build on this work (see, for example, the new Saturn
formation simulation by Dodson-Robinson et al. [2008]).  At Saturn's
heliocentric distance, 9.5~AU, the solid surface density never drops
below 8.7~g~cm$^{-2}$ during the first $1.5 \times 10^5$~yr of disk
evolution.  This large solid surface density should provide more than a
factor of 3 speedup in Saturn's solid growth phase over the Pollack et
al. results.

For Uranus, the relative increase in solids is even more pronounced: at
20~AU, the minimum solid surface density during the first $1.5 \times
10^5$~yr is 3.3~g~cm$^{-2}$, for a factor of 4 speedup in runaway core
growth.  Moving Uranus' starting position inward to 16~AU, in accordance
with the Nice model of planet migration (Tsiganis et al. 2005), provides
a starting $\Sigma_{\rm solid}$ of not less than 4.9~g~cm$^{-2}$, in
addition to the speedup given by a higher angular-frequency orbit.  The
prospects for Neptune to form rapidly at $\sim 12$~AU, as predicted by
the Nice model, are quite good: $\Sigma_{\rm solid} \ge
7.1$~g~cm$^{-2}$.

\cite{hubickyj05} provided an update to the Pollack et al. simulations
of Jupiter's growth, this time assuming rapid grain settling and
sublimation would lower opacities of the gaseous protoplanetary envelope
and speed up its contraction.  (Even if we assume low-opacity envelopes
for Saturn and Uranus, they still require high solid surface densities
to form quickly, as their solid growth stages both take $\ge 1.5$~Myr in
the Pollack et al. models.)  Hubickyj et al. found that, given a
planetesimal surface density of 10~g~cm$^{-2}$, Jupiter could reach its
present mass in 2.3~Myr.  Our work shows that this starting condition
for Jupiter is not only viable but likely: we predict solid surface
densities not less than 8.2~g~cm$^{-2}$ with refractory material only
and as high as 14.5~g~cm$^{-2}$ after the ice line moves through at $5
\times 10^4$~yr.

\section{Conclusions}
\label{conclusions}

By combining a viscously evolving protostellar disk model with a kinetic
model of ice formation, we have calculated the time-evolving solid
surface density available for giant planet formation.  We find three
results that are highly favorable for gas- and ice-giant formation:

\begin{enumerate}

\item The total (gas+solid) surface density distribution evolves toward
uniformity, with an outward-moving wave depositing mass in the giant
planet-forming region between 5 and 20~AU,

\item The ammonia and water ice lines coincide, providing a mass
enhancement of 7\% at the snow line over pure water condensation, and

\item Methane and CO condensation beyond 12~AU add mass to Uranus and
potentially Neptune's feeding zones.

\end{enumerate}

The next step is to use a core-accretion model to calculate how fast
Saturn, Uranus and Neptune can form with our enhanced surface densities.
The same methods used by \cite{hubickyj05} should work for Saturn.
However, an updated method of tracking planetesimal scattering may be
required for feeding zones with $R > 15$~AU, where orbital velocities of
planetesimals decrease to $\sim 10$--50 times their escape velocities
and self-stirring begins to affect the dynamics of the planetesimal
disk.  By calculating solid surface density as a function of radius and
time self-consistently with viscous evolution of the protostellar disk,
instead of assuming a single gas/solid ratio covers the entire giant
planet-forming region of the solar nebula, the research presented here
allows core accretion simulations to be fully deterministic.

One area that requires more work is the viscous evolution of an
MRI-turbulent disk with a dead zone, where $\alpha$ varies with radius.
Annuli with an inactive midplane would have a lower column-averaged
value of $\alpha$, which could slow or stop the movement of mass from
the inner solar nebula to the giant planet-forming region.  However,
Turner and Sano (2008) find that a toroidal magnetic field component
produced in the midplane by shear drives a laminar accretion flow even
through the nonturbulent dead zone.

Since our focus is on increasing the solid mass available for giant
planet cores, we have neglected clathrate hydrates in our chemical
model, preferring to tap into the large mass reservoir provided by ices.
However, noble gas enrichment in the giant planet atmospheres suggests
that clathration is very important, if not for building planet cores,
then at least for determining atmospheric composition during the late
stages of accretion (Hersant et al. 2004).  The presence of PAHs in
carbonaceous meteorites, cometary dust and Iapetus implies that these,
too, might be an important solid mass reservoir.  It is not yet known
whether PAHs survive in the ISM or form primarily in the warm disk.
Woods and Willacy (2007) present a mechanism for benzene formation in
the inner 3~AU of the solar nebula.  We are investigating the
possibility of including PAH chemistry in future work.

Finally, we note that the ALMA Design Reference Science Plan contains a
proposal for detecting the snow line in nearby protoplanetary disks.  We
submit that the CO ice line would be an equally valuable detection,
since this work strongly suggests that CO ice should be part of the
solid inventory of the Uranus- and Neptune-building planetesimals.  In a
1 Myr-old solar nebula analog in Taurus (140 pc), the CO ice line would
lie 6 resolution units from the star when observing the CO ($J = 3
\rightarrow 2$) transition.

We thank Julie Moses, Mark Marley and Doug Lin for helpful discussions
about the project design.  S.D.R. was supported by grants from the NSF
Graduate Research Fellowship program and the Achievement Rewards for
College Scientists Foundation.  K.W. and N.T. were supported by the JPL
Research and Technology Development Program.  P.B. received funding from
NSF Grant AST-0507424 and NASA Origins Grant NNX08AH82G.


\begin{deluxetable}{clcl}
\renewcommand{\baselinestretch}{1}
\tablecaption{Parameters Specifying Initial Solar Nebula
\label{freedisk} }
\tablehead{
\colhead{Parameter} & \colhead{Description} & \colhead{Value} &
\colhead{Units} }
\startdata
$M_*$	& star mass & 0.95	& $M_{\odot}$ \\
$M_{\rm disk}$ 	& disk mass & 0.12 & $M_{\odot}$ \\
$R_{\rm in}$	& inner radius & 0.3 & AU \\
$R_{\rm out}$	& outer radius & 30 & AU \\
$\Sigma(R)$ & mass distribution & $\Sigma \propto R^{-1.5}$ &
g~cm$^{-2}$ \\
$\alpha$	& viscosity parameter & 0.002 & dimensionless \\
$T_{\rm amb}$	& ambient temperature & 20 & K \\
\enddata
\renewcommand{\baselinestretch}{2}
\end{deluxetable}

\begin{deluxetable}{lrl}
\renewcommand{\baselinestretch}{1}
\tablecaption{Solar Composition Used for Opacity Table and Molecular
Cloud Model
\label{solarcomp} }
\tablehead{ \colhead{Element} & \colhead{Abundance\tablenotemark{a}} &
\colhead{Solids Formed} }
\startdata
H	&	12.00 & ice, refractory CHON \\
He	&	10.99 & none\\
C	&	8.60 & ice, refractory CHON \\
N	&	7.97 & ice, refractory CHON \\
O	&	8.87 & rock, ice, refractory CHON \\
Ne	&	8.07 & none \\
Mg	&	7.58 & rock \\
Si	&	7.55 & rock \\
S	&	7.21 & rock, ice \\
Fe	&	7.51 & metal \\
\enddata
\tablenotetext{a}{Abundances are logarithmic, with $\log \; n_H \equiv
12.00$}
\renewcommand{\baselinestretch}{2}
\end{deluxetable}

\begin{deluxetable}{clcl}
\renewcommand{\baselinestretch}{1}
\tabletypesize{\scriptsize}
\tablecaption{Free Parameters in Chemical Model of Disk Midplane
\label{freechem} }
\tablehead{
\colhead{Parameter} & \colhead{Description}  & \colhead{value} &
\colhead{units} }
\startdata
$\Sigma_{cr}$ & cosmic ray penetration depth & 150 & g~cm$^{-2}$ \\
$n_s$ & surface density of reaction sites on grains & $10^{15}$ &
cm$^{-2}$ \\
$a$ & mean grain radius & 0.258 & $\micron$ \\
$n_g$ & grain number density & $10^{-12} \times n_{\rm H}$ & cm$^{-3}$ \\
$S_x$ & sticking coefficient of species x & \( \left \{ \begin{array}{l}
0.3, \;  x = {\rm H}  \\ 1.0, \;  x \ne {\rm H} \end{array} \right. \) 
& dimensionless \\
${\rm n_H / (n_H + 2n_{H2})} $ & atomic hydrogen fraction & 0.01 & dimensionless \\
$n_{\rm C, grain} / n_{\rm C}$ & fraction of carbon in grains & 0.4 &
dimensionless \\
$n_{\rm N, grain} / n_{\rm N}$ & fraction of nitrogen in grains & 0.1 &
dimensionless \\
$n_{\rm O, grain} / n_{\rm O}$ & fraction of oxygen in grains & 0.2 &
dimensionless \\
$n_{\rm S, grain} / n_{\rm S}$ & fraction of sulfur in grains & 0.85 &
dimensionless \\
\enddata
\renewcommand{\baselinestretch}{2}
\end{deluxetable}

\begin{deluxetable}{lrc}
\renewcommand{\baselinestretch}{1}
\tabletypesize{\scriptsize}
\tablecaption{Binding energies used to determine thermal desorption
rates of ices
\label {binding}}
\tablewidth{0pt}
\tablehead{ \colhead{Species} & \colhead{Binding Energy (K)} &
\colhead{Reference\tablenotemark{a}} \\ }
\startdata
H	&	600\tablenotemark{b}	&	CT	\\
C	&	800	&	TA	\\
N	&	800	&	TA	\\
O	&	800	&	TA	\\
S	&	1100	&	TA	\\
N$_2$	&	790	&	O	\\
NO	&	1210	&	HH	\\
CO	&	1720\tablenotemark{b}	&	SA	\\
H$_2$O	&	5700	& 	C	\\
HCN	&	1760	&	HH	\\
H$_2$S	&	1800	& 	HH	\\
CO$_2$	&	2860	&	SA	\\
OCS	&	3000	&	HH	\\
NH$_3$	&	5700	& 	C	\\
CH$_4$	&	1360	&	HH	\\
H$_2$CS	&	2250	&	HH	\\
C$_2$H$_2$	&	1610	&	HH	\\
\enddata
\tablenotetext{a}{
References: CT = Cazaux \& Tielens (2002); TA = Tielens \& Allamandola
(1987); O = \"{O}berg et al. (2005); HH = Hasegawa \& Herbst (1993); SA
= Sandford \& Allamandola (1990); C = Collings et al. (2004).}
\tablenotetext{b}{Surface binding energy on an amorphous H$_2$O ice
substrate.}
\renewcommand{\baselinestretch}{2}
\end{deluxetable}

\begin{deluxetable}{lr}
\renewcommand{\baselinestretch}{1}
\tabletypesize{\scriptsize}
\tablecaption{Output Abundances from Molecular Cloud Simulation
\label{cloudoutput}} 
\tablehead{
\colhead{Species} & \colhead{Abundance} }
\startdata
H	&	4.08 (-05)\tablenotemark{a} \\
He	&	9.77 (-02) \\
C	&	2.61 (-11) \\
N	&	2.65 (-11) \\
O	&	3.76 (-11) \\
H$^+$	&	4.48 (-07) \\
He$^+$	&	7.01 (-09) \\
H$_3^+$	&	2.57 (-10) \\
H$_2$	&	4.99 (-01) \\
CS	&	8.07 (-15) \\
GCO\tablenotemark{b}	&	2.82 (-05) \\
GSi	&	2.00 (-08) \\
GNO	&	7.43 (-07) \\
GH$_2$O	&	5.27 (-04) \\
GHCN	&	7.20 (-06) \\
GHNC	&	8.74 (-08) \\
GCO$_2$	&	3.80 (-09) \\
GNH$_3$	&	7.56 (-05) \\
GC$_2$H$_2$	&	1.48 (-07) \\
GH$_2$CO	&	9.80 (-08) \\
GCH$_4$	&	2.02 (-04) \\
GC$_3$H$_2$	&	2.94 (-07) \\
GHC$_3$N	&	1.31 (-08) \\
GCH$_3$OH	&	1.96 (-09) \\
GCH$_3$CN	&	4.53 (-09) \\
GN$_2$	&	1.85 (-07) \\
GCH$_2$CO	&	9.80 (-10) \\
GC$_2$O	&	1.12 (-10) \\
GCH$_3$CHO	&	1.61 (-11) \\
GC$_2$H$_4$	&	2.38 (-08) \\
GH$_2$S	&	2.42 (-06) \\
GH$_2$CS	&	2.54 (-09) \\
GC$_2$S	&	6.60 (-09) \\
GSO$_2$	&	5.06 (-10) \\
GC$_3$S	&	5.19 (-10) \\
GC$_2$H$_6$	&	3.77 (-12) \\
GC$_3$H$_4$	&	7.14 (-14) \\
\enddata
\tablenotetext{a}{All abundances are fractional with respect to the
total number of hydrogen nuclei, $n_H + 2n_{H2}$.  1.00 (-10) should be
read as $1 \times 10^{-10}$.}
\tablenotetext{b}{G denotes solid-phase grain mantle species.}
\renewcommand{\baselinestretch}{2}
\end{deluxetable}

\begin{deluxetable}{ll}
\renewcommand{\baselinestretch}{1}
\tablecaption{Abundance Ratios of Major Ices After Molecular Cloud
Phase
\label{iceratios} }
\tablehead{ \colhead{Ice Species} & \colhead{$n_x / n_{H_2O}$}
}
\startdata
H$_2$O & $\equiv 1$ \\
CH$_4$ & 0.38 \\
NH$_3$ & 0.14 \\
CO & 0.054 \\
HCN & 0.014 \\
NO & 0.0014 \\
N$_2$ & 0.00035 \\
\enddata
\renewcommand{\baselinestretch}{2}
\end{deluxetable}

\begin{deluxetable}{rrrrrrrrrrrrrrrr}
\renewcommand{\baselinestretch}{1}
\rotate
\tabletypesize{\tiny}
\tablecaption{Solid Surface Density as a Function of Radius and Time
\label{sdtable} }
\tablehead{
\colhead{} & \multicolumn{15}{c}{Radius (AU)} \\
\colhead{Time (Myr)} &
 \colhead{1} & \colhead{2} &  \colhead{3} &  \colhead{4} & \colhead{5} &
\colhead{6} & \colhead{7} & \colhead{8} & \colhead{9} & \colhead{10} &
\colhead{12} & \colhead{14} & \colhead{16} & \colhead{18} &
\colhead{20} }
\startdata
  0 & 110.63\tablenotemark{a} &  39.11 &  21.29 &  13.83 &   9.90 &  16.28 &  13.04 &  10.68 &   9.15 &   8.23 &   7.07 &   5.61 &   4.59 &   3.85 &   3.28  \\
  0.01 &  38.48 &  24.14 &  17.25 &  12.73 &   9.94 &  12.62 &  15.11 &  12.58 &  10.28 &   8.85 &   7.31 &   5.72 &   4.64 &   3.87 &   3.29  \\
  0.02 &  25.95 &  18.06 &  13.51 &  10.70 &   8.94 &  16.08 &  15.02 &  13.07 &  11.11 &   9.61 &   7.81 &   5.88 &   4.70 &   3.89 &   3.30  \\
  0.03 &  20.70 &  14.88 &  11.39 &   9.37 &   8.22 &  15.94 &  14.41 &  12.85 &  11.24 &   9.85 &   8.34 &   6.13 &   4.79 &   3.93 &   3.31  \\
  0.04 &  17.57 &  12.83 &  10.02 &   8.46 &   8.68 &  15.07 &  13.78 &  12.48 &  11.11 &   9.90 &   8.67 &   6.41 &   4.92 &   3.97 &   3.32  \\
  0.05 &  15.42 &  11.37 &   9.06 &   7.81 &  12.77 &  14.33 &  13.22 &  12.08 &  10.88 &   9.86 &   8.84 &   6.66 &   5.08 &   4.03 &   3.33  \\
  0.06 &  13.90 &  10.29 &   8.34 &   7.33 &  14.28 &  13.70 &  12.73 &  11.70 &  10.63 &   9.78 &   8.90 &   6.85 &   5.23 &   4.10 &   3.35  \\
  0.07 &  12.75 &   9.47 &   7.79 &   7.04 &  14.04 &  13.17 &  12.33 &  11.36 &  10.38 &   9.68 &   8.91 &   6.98 &   5.37 &   4.18 &   3.36  \\
  0.08 &  11.84 &   8.84 &   7.36 &   7.15 &  13.54 &  12.73 &  11.97 &  11.05 &  10.15 &   9.58 &   8.89 &   7.07 &   5.49 &   4.26 &   3.39  \\
  0.09 &  11.10 &   8.33 &   7.01 &   8.27 &  13.09 &  12.34 &  11.65 &  10.77 &   9.94 &   9.46 &   8.86 &   7.12 &   5.57 &   4.32 &   3.41  \\
  0.1 &  10.49 &   7.91 &   6.72 &  10.40 &  12.70 &  12.00 &  11.35 &  10.52 &   9.75 &   9.34 &   8.81 &   7.14 &   5.64 &   4.38 &   3.44  \\
  0.15 &   8.53 &   6.59 &   5.85 &  11.83 &  11.35 &  10.83 &  10.24 &   9.56 &   9.10 &   8.72 &   8.46 &   7.05 &   5.74 &   4.55 &   3.54  \\
  0.2 &   7.46 &   5.89 &   6.19 &  10.91 &  10.50 &  10.05 &   9.51 &   8.93 &   8.70 &   8.28 &   8.07 &   6.82 &   5.66 &   4.55 &   3.57  \\
  0.3 &   6.28 &   5.12 &  10.03 &   9.80 &   9.48 &   9.04 &   8.56 &   8.28 &   7.95 &   8.04 &   7.41 &   6.35 &   5.35 &   4.37 &   3.49  \\
  0.4 &   5.62 &   4.69 &   9.51 &   9.11 &   8.80 &   8.37 &   7.96 &   7.78 &   7.54 &   7.80 &   6.88 &   5.94 &   5.03 &   4.14 &   3.35  \\
  0.5 &   5.17 &   4.43 &   8.99 &   8.59 &   8.29 &   7.87 &   7.60 &   7.34 &   7.46 &   7.39 &   6.46 &   5.59 &   4.74 &   3.92 &   3.20  \\
  0.6 &   4.84 &   4.33 &   8.57 &   8.19 &   7.88 &   7.48 &   7.31 &   7.06 &   7.33 &   7.00 &   6.11 &   5.29 &   4.48 &   3.72 &   3.05  \\
  0.7 &   4.58 &   4.65 &   8.23 &   7.86 &   7.55 &   7.18 &   7.02 &   6.94 &   7.08 &   6.67 &   5.82 &   5.03 &   4.25 &   3.54 &   2.92  \\
  0.8 &   4.37 &   5.71 &   7.94 &   7.57 &   7.27 &   6.95 &   6.75 &   6.91 &   6.80 &   6.38 &   5.56 &   4.80 &   4.05 &   3.38 &   2.79  \\
  0.9 &   4.19 &   7.03 &   7.69 &   7.32 &   7.02 &   6.78 &   6.54 &   6.83 &   6.55 &   6.12 &   5.34 &   4.59 &   3.87 &   3.23 &   2.68  \\
  1 &   4.03 &   7.56 &   7.46 &   7.10 &   6.79 &   6.62 &   6.39 &   6.68 &   6.31 &   5.90 &   5.13 &   4.40 &   3.71 &   3.10 &   2.58  \\
\enddata
\tablenotetext{a}{Surface density units are g~cm$^{-2}$.}
\renewcommand{\baselinestretch}{2}
\end{deluxetable}

\clearpage



\begin{thebibliography}{}

\bibitem[Abe et al.(2000)]{abe00} Abe, Y., Ohtani, E., Okuchi, T.,
Righter, K., Draker, M.\ 2000.\ Water in the Early Earth.\ Origin of the
earth and moon, edited by R.M.~Canup and K.~Righter and 69 collaborating
authors. Tucson: University of Arizona Press, p. 413-433

\bibitem[A'Hearn(2008)]{ahearn08} A'Hearn, M.~F.\ 2008.\ Deep Impact and
the Origin and Evolution of Cometary Nuclei.\ Space Science Reviews 56.

\bibitem[Aikawa et al.(1996)]{aikawa96} Aikawa, Y., Miyama, S.~M.,
Nakano, T., Umebayashi, T.\ 1996.\ Evolution of Molecular Abundance in
Gaseous Disks around Young Stars: Depletion of CO Molecules.\
Astrophysical Journal 467, 684.

\bibitem[Aikawa et al.(2008)]{aikawa08} Aikawa, Y., Wakelam, V., Garrod,
R.~T., Herbst, E.\ 2008.\ Molecular Evolution and Star Formation: From
Prestellar Cores to Protostellar Cores.\ Astrophysical Journal 674, 984.

\bibitem[Alexander et al.(2006)]{alexander06} Alexander, R.~D., Clarke,
C.~J., Pringle, J.~E.\ 2006.\ Photoevaporation of protoplanetary
discs - II. Evolutionary models and observable properties.\ Monthly
Notices of the Royal Astronomical Society 369, 229-239.

\bibitem[Allamandola and Sandford(1994)]{allamandola94} Allamandola,
L.~J., Sandford, S.~A.\ 1994.\ The First Symposium on the Infrared
Cirrus and Diffuse Interstellar Clouds, 58, 302

\bibitem[Allen and Robinson(1997)]{allen77} Allen, M., Robinson, G.~W.
1977. The molecular composition of dense interstellar clouds.\
Astrophysical Journal 212, 396-415.

\bibitem[Anders and Grevesse(1989)]{anders89} Anders, E., Grevesse, N.
1989.\ Abundances of the elements - Meteoritic and solar.\ Geochimica et
Cosmochimica Acta 53, 197-214

\bibitem[Balbus and Hawley(1991)]{balbus91} Balbus, S.~A., Hawley,
J.~F. 1991.\ A powerful local shear instability in weakly magnetized
disks. I - Linear analysis. II - Nonlinear evolution.\ Astrophysical
Journal 376, 214-233.

\bibitem[Barucci et al.(1996)]{barucci96} Barucci, M.~A., Fulchignoni,
M., Lazzarin, M.\ 1996.\ Water ice in primitive asteroids? Planetary and
Space Science 44, 1047-1049.

\bibitem[Barzel and Biham(2007)]{barzel07} Barzel, B., Biham, O. 2007.\
Efficient stochastic simulations of complex reaction networks on
surfaces.\ Journal of Chemical Physics 127, 4703.

\bibitem[Bell et al.(1997)]{bell97} Bell, K. R., Cassen, P. M., Klahr,
H. H., Henning, Th. 1997.\ The Structure and Appearance of Protostellar
Accretion Disks: Limits on Disk Flaring.\ Astrophysical Journal 486,
372.


\bibitem[Boss(2005)]{boss05} Boss, A.~P. 2005.\ Evolution of the Solar
Nebula. VII. Formation and Survival of Protoplanets Formed by Disk
Instability.\ Astrophysical Journal 629, 535-548.


\bibitem[Caselli et al.(1998)]{caselli98} Caselli, P., Hasegawa, T.~I.,
Herbst, E. 1998.\ A Proposed Modification of the Rate Equations for
Reactions on Grain Surfaces.\ Astrophysical Journal 495, 309.

\bibitem[Cazaux and Tielens(2002)]{cazaux02} Cazaux, S., Tielens,
A.~G.~G.~M. 2002. Molecular Hydrogen Formation in the Interstellar
Medium. Astrophysical Journal 575, L29-L32.

\bibitem[Chiang and Goldreich(1997)]{chiang97} Chiang, E.~I., Goldreich,
P.\ 1997.\ Spectral Energy Distributions of T Tauri Stars with Passive
Circumstellar Disks.\ Astrophysical Journal 490, 368.

\bibitem[Ciesla(2007)]{ciesla07} Ciesla, F.~J.\ 2007.\ Outward Transport
of High-Temperature Materials Arount the Midplane of the Solar Nebula.\
Science 318, 613.

\bibitem[Ciesla and Cuzzi(2006)]{ciesla06} Ciesla, F.~J., Cuzzi, J.~N.\
2006.\ The evolution of the water distribution in a viscous
protoplanetary disk.\ Icarus 181, 178-204.

\bibitem[Charnley and Rodgers(2008)]{charnley08} Charnley, S.~B., 
Rodgers, S.~D.\ 2008.\ Interstellar Reservoirs of Cometary Matter.\
Space Science Reviews 40.

\bibitem[Collings et al.(2004)]{collings04} Collings, M.~P., Anderson,
M.~A., Chen, R., Dever, J.~W., Viti, S., Williams, D.~A., McCoustra,
M.~R.~S.\ 2004.\ A laboratory survey of the thermal desorption of
astrophysically relevant molecules.\ Monthly Notices of the Royal
Astronomical Society 354, 1133-1140.

\bibitem[D'Alessio et al.(2006)]{dalessio06} D'Alessio, P., Calvet, N.,
Hartmann, L., Franco-Hern\'{a}ndez, R., Serv\'{i}n, H.\ 2006.\ Effects
of Dust Growth and Settling in T Tauri Disks. Astrophysical Journal
638, 314-335.

\bibitem[Desch(2007)]{desch07} Desch, S.~J.\ 2007.\ Mass Distribution
and Planet Formation in the Solar Nebula. Astrophysical Journal 671,
878-893.

\bibitem[Dodson-Robinson et al.(2008)]{dodsonrobinson08}
Dodson-Robinson, S.~E., Bodenheimer, P., Laughlin, G., Willacy, K.,
Turner, N.~J., Beichman, C.~A.\ 2008.\ Saturn Forms by Core Accretion in
3.4~Myr.\ ArXiv e-prints arXiv:0810.0228 (accepted for publication in
Astrophysical Journal Letters).

\bibitem[Dullemond et al.(2006)]{dullemond06} Dullemond, C.~P., Apai,
D., Walch, S.\ 2006.\ Crystalline Silicates as a Probe of Disk Formation
History.\ Astrophysical Journal 640, L67-L70.

\bibitem[Ferguson et al.(2005)]{ferguson05} Ferguson, J.~W., Alexander,
D.~R., Allard, F., Barman, T., Bodnarik, J.~G., Hauschildt, P.~T.,
Heffner-Wong, A., Tamanai, A.\ 2005.\ Low-Temperature Opacities.
Astrophysical Journal 623, 585-596.


\bibitem[Flasar et al.(2005)]{flasar05} Flasar, F.~M., and 45
colleagues. 2005.\ Temperatures, Winds, and Composition in the Saturnian
System.\ Science 307, 1247-1251.

\bibitem[Fraser et al.(2001)]{fraser01} Fraser, H.~J., Collings, M.~P.,
McCoustra, M.~R.~S., Williams, D.~A.\ 2001.\ Thermal desorption of water
ice in the interstellar medium. Monthly Notices of the Royal
Astronomical Society 327, 1165-1172

\bibitem[Freeman et al.(2007)]{freeman07} Freeman, J., Stegman, D.,
May, D.\ 2007.\ The Role of Ammonia in the Evolution of Enceladus.\ AGU
Fall Meeting Abstracts 7.

\bibitem[Garrod and Herbst(2006)]{garrod06} Garrod, R. T., Herbst, E.\
2006.\ Formation of methyl formate and other organic species in the
warm-up phase of hot molecular cores. Astronomy and Astrophysics 457,
927-936.

\bibitem[Grundy et al.(2002)]{grundy02} Grundy, W.~M., Buie, M.~W., 
Spencer, J.~R.\ 2003.\ Spectroscopy of Pluto and Triton at 3-4 Microns:
Possible Evidence for Wide Distribution of Nonvolatile Solids.
Astronomical Journal 124, 2273-2278.

\bibitem[Haisch et al.(2001)]{haisch01} Haisch, K.~E., Jr., Lada, E.~A.,
Lada, C.~J.\ 2001.\ Disk Frequencies and Lifetimes in Young Clusters.\
Astrophysical Journal 553, L153-L156.

\bibitem[Hallenbeck et al.(2000)]{hallenbeck00} Hallenbeck, S.~L., Nuth,
J.~A. III, Nelson, R.~N.\ 2000.\ Evolving Optical Properties of
Annealing Silicate Grains: From Amorphous Condensate to Crystalline
Mineral.\ Astrophysical Journal 535, 247-255.

\bibitem[Harker and Desch(2002)]{harker02} Harker, D.~E., Desch, S.~J.\
2002.\ Annealing of Silicate Dust by Nebular Shocks at 10~AU.
Astrophysical Journal 565, L109-L112.

\bibitem[Hasegawa and Herbst(1993)]{hasegawa93} Hasegawa, T.~I., 
Herbst, E.\ 1993.\ New gas-grain chemical models of quiescent dense
interstellar clouds - The effects of H2 tunnelling reactions and cosmic
ray induced desorption. Monthly Notices of the Royal Astronomical
Society 261, 83-102

\bibitem[Hasegawa et al.(1992)]{hasegawa92} Hasegawa, T.~I., Herbst, E.,
Leung, C.~M.\ 1992.\ Models of gas-grain chemistry in dense interstellar
clouds with complex organic molecules.\ Astrophysical Journal Supplement
Series 82, 167-195.

\bibitem[Hawley and Stone(1998)]{hawley98} Hawley, J.~F., Stone, J.~M.\
1998.\ Nonlinear Evolution of the Magnetorotational Instability in
Ion-Neutral Disks.\ Astrophysical Journal 501, 758.

\bibitem[Hawley et al.(1999)]{hawley99} Hawley, J.~F., Balbus, S.~A., 
Winters, W.~F.\ 1999.\ Local Hydrodynamic Stability of Accretion Disks.\
Astrophysical Journal 518, 394-404.

\bibitem[Hayashi (1981)]{hayashi81} Hayashi, C.\ 1981.\ Structure of the
Solar Nebula, Growth and Decay of Magnetic Fields and Effects of
Magnetic and Turbulent Viscosities on the Nebula.\ Progress of
Theoretical Physics Supplement 70, 35-53.

\bibitem[Helling et al.(2000)]{helling00} Helling, Ch., Winters, J.~M.,
Sedlmayr, E.\ 2000.\ Circumstellar dust shells around long-period
variables. VII. The role of molecular opacities.\ Astronomy and
Astrophysics 358, 651-664.

\bibitem[Herbst et al.(2005)]{herbst05} Herbst, E., Chang, Q., Cuppen,
H.~M.\ 2005.\ Chemistry on interstellar grains.\ Journal of Physics
Conference Series, 6, 18-35.

\bibitem[Hersant et al.(2001)]{hersant01} Hersant, F., Gautier, D., 
Hur\'{e}, J.-M.\ 2001.\ A Two-dimensional Model for the Primordial
Nebula Constrained by D/H Measurements in the Solar System: Implications
for the Formation of Giant Planets.\ Astrophysical Journal 554, 391-407.

\bibitem[Hersant et al.(2004)]{hersant04} Hersant, F., Gautier, D., 
Lunine, J.~I.\ 2004.\ Enrichment in volatiles in the giant planets of
the Solar System.\ Planetary and Space Science 52, 623-641.


\bibitem[Hornekaer et al.(2003)]{hornekaer03} Hornekaer, L., Baurichter,
A., Petrunin, V.~V., Field, D., Luntz, A.~C.\ 2003.\ Importance of
Surface Morphology in Interstellar H$_2$ Formation.\ Science 302,
1943-1946.

\bibitem[Hubbard and Blackman(2006)]{hubbard06} Hubbard, A., Blackman,
E.~G.\ 2006.\ Planetesimal growth in turbulent discs before the onset of
gravitational instability.\ New Astronomy 12, 246-263.

\bibitem[Hubickyj et al.(2005)]{hubickyj05} Hubickyj, O., Bodenheimer,
P., Lissauer, J.~J.\ 2005.\ Accretion of the gaseous envelope of Jupiter
around a 5-10 Earth-mass core.\ Icarus 179, 415-431.

\bibitem[Hussmann et al.(2006)]{hussmann06} Hussmann, H., Sohl, F., 
Spohn, T.\ 2006.\ Subsurface oceans and deep interiors of medium-sized
outer planet satellites and large trans-neptunian objects. Icarus 185,
258-273.

\bibitem[Ilgner and Nelson(2006)]{ilgner06} Ilgner, M., Nelson, R.~P.\
2006.\ On the ionisation fraction in protoplanetary disks. II. The
effect of turbulent mixing on gas-phase chemistry.\ Astronomy and
Astrophysics 445, 223-232.

\bibitem[Jessberger et al.(1988)]{jessberger88} Jessberger, E. K.,
Christoforidis, A., Kissel, J.\ 1988.\ Aspects of the major element
composition of Halley's dust.\ Nature 332, 691-695.

\bibitem[Keller et al.(2006)]{keller06} Keller, L.~P., and 32
colleagues. 2006.\ Infrared Spectroscopy of Comet 81P/Wild 2 Samples
Returned by Stardust.\ Science 314, 1728.

\bibitem[Kippenhahn and Weigert(1994)]{kippenhahn94} Kippenhahn, R.,
Weigert, A.\ 1994, Stellar Structure and Evolution (Berlin:
Springer-Verlag)

\bibitem[Kretke and Lin(2007)]{kretke07} Kretke, K.~A.,  Lin, D.~N.~C.\
2007.\ Grain Retention and Formation of Planetesimals near the Snow Line
in MRI-driven Turbulent Protoplanetary Disks.\ Astrophysical Journal
664, L55-L58.

\bibitem[Lodders(2003)]{lodders03} Lodders, K.\ 2003.\ Solar System
Abundances and Condensation Temperatures of the Elements.\ Astrophysical
Journal 591, 1220-1247.

\bibitem[Lodders(2004)]{lodders04} Lodders, K.\ 2004.\ Jupiter Formed
with More Tar than Ice.\ Astrophysical Journal 611, 587-597.

\bibitem[Lodders and Fegley(1994)]{lodders94} Lodders, K., Fegley, B.\
1994.\ The origin of carbon monoxide in Neptune's atmosphere.\ Icarus,
112, 368-375.

\bibitem[Lopes et al.(2007)]{lopes07} Lopes, R.~M.~C., and 43
colleagues 2007.\ Cryovolcanic features on Titan's surface as revealed
by the Cassini Titan Radar Mapper.\ Icarus 186, 395-412.

\bibitem[Lynden-Bell and Pringle(1974)]{lyndenbell74} Lynden-Bell, D.,
Pringle, J.~E.\ 1974.\ The evolution of viscous discs and the origin of
the nebular variables.\ Monthly Notices of the Royal Astronomical
Society 168, 603-637.


\bibitem[Lyra et al.(2008)]{lyra08} Lyra, W., Johansen, A., Klahr, H.,
and Piskunov, N.\ 2008.\ Global magnetohydrodynamical models of
turbulence in protoplanetary disks. I. A cylindrical potential on a
Cartesian grid and transport of solids.\ Astronomy and Astrophysics 479,
883-901.

\bibitem[Mihalas (1978)]{mihalas78} Mihalas, D. 1978, Stellar
Atmospheres (San Francisco: Freeman)

\bibitem[Millar et al.(1997)]{millar97} Millar, T.~J., Farquhar, P.~R.
A., Willacy, K.\ 1997.\ The UMIST Database for Astrochemistry 1995.\
Astronomy and Astrophysics Supplement Series 121, 139-185.

\bibitem[Milsom et al.(1994)]{milsom94} Milsom, J. A., Chen, X., 
Taam, R.\ 1994.\ The vertical structure and stability of accretion disks
surrounding black holes and neutron stars.\ Astrophysical Journal 421,
668-676.

\bibitem[\"{O}berg et al.(2005)]{oberg05} \"{O}berg, K.~I., van
Broekhuizen, F., Fraser, H.~J., Bisschop, S.~E., van Dishoeck, E.~F., 
Schlemmer, S.\ 2005.\ Competition between CO and N$_2$ Desorption from
Interstellar Ices.\ Astrophysical Journal 631, L33-L36.

\bibitem[Owen(2000)]{owen00} Owen, T.\ 2000.\ On the origin of Titan's
atmosphere. Planetary and Space Science 48, 747-752.

\bibitem[Owen and Encrenaz(2003)]{owen03} Owen, T., Encrenaz, T.\ 2003.\
Element Abundances and Isotope Ratios in the Giant Planets and Titan.\
Space Science Reviews 106, 121-138.

\bibitem[Podolak(2003)]{podolak03} Podolak, M.\ 2003.\ The contribution
of small grains to the opacity of protoplanetary atmospheres.\ Icarus
165, 428-437.

\bibitem[Pollack et al.(1996)]{pollack96} Pollack, J.~B., Hubickyj, O.,
Bodenheimer, P., Lissauer, J.~J., Podolak, M., Greenzweig, Y.\ 1996.\
Formation of the Giant Planets by Concurrent Accretion of Solids and
Gas.\ Icarus 124, 62-85.

\bibitem[Prentice(2007)]{prentice07} Prentice, A.~J.\ 2007.\ Iapetus: a
Prediction for Bulk Chemical Composition, Internal Physical Structure
and Origin.\ AGU Fall Meeting Abstracts 1432.

\bibitem[Pringle(1981)]{pringle81} Pringle, J.~E.\ 1981.\ Accretion
discs in astrophysics.\ Annual Review of Astronomy and Astrophysics 19,
137-162.

\bibitem[Press et al.(1992)]{press92} Press, W. H., Teukolsky, S. A.,
Vetterling, W. T., Flannery, B. P. 1992, Numerical Recipes in Fortran
77, Second Edition (Cambridge: Cambridge University Press)

\bibitem[Reyes-Ruiz et al.(2003)]{reyesruiz03} Reyes-Ruiz, M.,
P\'{e}rez-Tijerina, E., S\'{a}nchez-Salcedo, F.~J.\ 2003.\ The
Magneto-Rotational Instability in Protoplanetary Disks.\ Revista
Mexicana de Astronomia y Astrofisica Conference Series 18, 92-96.


\bibitem[Rivkin et al.(2000)]{rivkin00} Rivkin, A.~S., Howell, E.~S.,
Vilas, F., Lebofsky, L.~A.\ 2002.\ Hydrated Minerals on Asteroids: The
Astronomical Record.\ Asteroids III 235-253.


\bibitem[Ruffle and Herbst(2000)]{ruffle00} Ruffle, D.~P., Herbst, E.\
2000.\ New models of interstellar gas-grain chemistry - I. Surface
diffusion rates.\ Monthly Notices of the Royal Astronomical Society 319,
837-850.

\bibitem[Ruden and Lin(1986)]{ruden86} Ruden, S.~P., Lin, D.~N.~C.\
1986.\ The global evolution of the primordial solar nebula.\
Astrophysical Journal 308, 883-901.

\bibitem[Ruden and Pollack(1991)]{ruden91} Ruden, S.~P., Pollack,
J.~B.\ 1991.\ The dynamical evolution of the protosolar nebula.\
Astrophysical Journal 375, 740-760.

\bibitem[Safronov(1969)]{safronov69} Safronov, V.~S. 1969, In Russian,
English translation: NASA-TTF-677, 1972

\bibitem[Sandford and Allamandola(1990)]{sandford90} {Sandford, S.~A., 
Allamandola, L.~J.\ 1990.\ The volume- and surface-binding energies of
ice systems containing CO, CO2, and H2O.\ Icarus 87, 188-192.


\bibitem[Semenov et al.(2003)]{semenov03} Semenov, D., Henning, Th.,
Helling, Ch., Ilgner, M., Sedlmayr, E.\ 2003.\ Rosseland and Planck mean
opacities for protoplanetary discs.\ Astronomy and Astrophysics 410,
611-621.

\bibitem[Shakura and Syunyaev(1973)]{shakura73} Shakura, N.~I., 
Syunyaev, R.~A.\ 1973.\ Black holes in binary systems. Observational
appearance.\ Astronomy and Astrophysics 24, 337-355

\bibitem[Spohn and Schubert(2003)]{spohn03} Spohn, T., Schubert, G.\
2003.\ Oceans in the icy Galilean satellites of Jupiter? Icarus 161,
456-467.

\bibitem[Stamatellos and Whitworth(2008)]{stamatellos08} Stamatellos,
D., Whitworth, A.~P.\ 2008.\ Can giant planets form by gravitational
fragmentation of discs? Astronomy and Astrophysics 480, 879-887.

\bibitem[Stepinski and Valageas(1997)]{stepinski97} Stepinski, T.~F.,
Valageas, P.\ 1997.\ Global evolution of solid matter in turbulent
protoplanetary disks. II. Development of icy planetesimals.\ Astronomy
and Astrophysics 319, 1007-1019.

\bibitem[Tielens and Allamandola(1987)]{tielens87} Tielens, A.~G.~G.~M.,
Allamandola, L.~J.\ 1987.\ Composition, structure, and chemistry of
interstellar dust.\ Interstellar Processes 134, 397-469.

\bibitem[Tielens and Hagen(1982)]{tielens82} Tielens, A.~G.~G.~M., 
Hagen, W.\ 1982.\ Model calculations of the molecular composition of
interstellar grain mantles.\ Astronomy and Astrophysics 114, 245-260.


\bibitem[Tsiganis et al.(2005)]{tsiganis05} Tsiganis, K., Gomes, R.,
Morbidelli, A., Levison, H.~F.\ 2005.\ Origin of the orbital
architecture of the giant planets of the Solar System. Nature 435,
459-461.

\bibitem[Turner and Sano(2008)]{turner08} Turner, N.~J., Sano, T.\
2008.\ Dead Zone Accretion Flows in Protostellar Disks.\ Astrophysical
Journal 679, L131-L134.


\bibitem[Umebayashi and Nakano(1980)]{umebayashi80} Umebayashi, T., 
Nakano, T.\ 1980.\ Recombination of Ions and Electrons on Grains and the
Ionization Degree in Dense Interstellar Clouds.\ Publications of the
Astronomical Society of Japan 32, 405.

\bibitem[Umebayashi and Nakano(1981)]{umebayashi81} Umebayashi, T., 
Nakano, T.\ 1981.\ Fluxes of Energetic Particles and the Ionization Rate
in Very Dense Interstellar Clouds.\ Publications of the Astronomical
Society of Japan 33, 617.

\bibitem[van Boekel et al.(2004)]{vanboekel04} van Boekel, R., and 22
colleagues. 2004.\ The building blocks of planets within the
`terrestrial' region of protoplanetary disks.\ Nature 432, 479-482.

\bibitem[Weidenschilling(1977)]{weidenschilling77} Weidenschilling,
S.~J.\ 1977.\ The distribution of mass in the planetary system and solar
nebula.\ Astrophysics and Space Science 51, 153-158.


\bibitem[Willacy and Langer(2000)]{willacy00} Willacy, K., Langer,
W.~D.\ 2000.\ The Importance of Photoprocessing in Protoplanetary
Disks.\ Astrophysical Journal 544, 903-920.

\bibitem[Willacy et al.(2006)]{willacy06} Willacy, K., Langer, W.,
Allen, M., Bryden, G.\ 2006.\ Turbulence-driven Diffusion in
Protoplanetary Disks: Chemical Effects in the Outer Regions.\
Astrophysical Journal 644, 1202-1213.

\bibitem[Woods and Willacy(2007)]{woods07} Woods, P.~M., Willacy, K.
2007.\ Benzene Formation in the Inner Regions of Protostellar Disks.\
Astrophysical Journal 655, L49-L52.

\bibitem[Womack et al.(1992)]{womack92} Womack, M., Wyckoff, S., 
Ziurys, L.~M.\ 1992.\ Estimates of N2 abundances in dense molecular
clouds.\ Astrophysical Journal 401, 728-735.

\end{thebibliography}
\end{document}